\documentclass[12pt,a4paper]{article} 

\usepackage{amsmath,amssymb,cite}
\usepackage{epsfig}

\def\nslash{\rlap{\hspace{0.02cm}/}{n}}

\def\vslash{\rlap{\hspace{0.02cm}/}{v}}

\def\calAslash{\rlap{\hspace{0.08cm}/}{{\cal A}}}
\def\epsslash{\rlap{\hspace{0.02cm}/}{\varepsilon}}

\def\etaslash{\rlap{\hspace{0.02cm}/}{\eta}}
\newcommand{\MSbar}{\overline{\rm MS}}
\newcommand{\eps}{\epsilon}
\newcommand{\Unew}{u}
\begin{document} 

\begin{titlepage}

\begin{flushright}
DESY 07-124\\
September 2007 \\  
\end{flushright}

\vspace*{5mm}
\begin{center}
    {\baselineskip 25pt
    \Large\bf

\boldmath{Towards  $B\to V\gamma$ Decays at NNLO in SCET} 
    }

\vspace{1.2cm}
\centerline{\bf 
Ahmed Ali\footnote{E-mail: ahmed.ali@desy.de},
Ben D. Pecjak\footnote{E-mail: pecjak@mail.desy.de}
}

\vspace{.5cm}
\small{\it Theory Group,
 Deutsches Elektronen-Synchrotron DESY,
 D-22603 Hamburg, Germany.}
\vspace{.5cm}

\centerline{\bf Christoph Greub\footnote{Email: greub@itp.unibe.ch}}

\vspace{.5cm}
\small{\it Institute for Theoretical Physics, Univ. Berne, CH-3012 Berne,
Switzerland.}

\end{center}

\centerline{\today} 

\bigskip

\centerline{\bf Abstract}

\medskip 

\noindent We compute NNLO (${\cal O}(\alpha_s^2)$) 
corrections to the hard-scattering kernels entering
the QCD factorization formula for $B\to V\gamma$ decays, 
where $V$  is a light vector meson. 
We give complete NNLO results for the dipole operators 
$Q_7$ and $Q_8$, and partial results for $Q_1$ valid in  
the large-$\beta_0$ limit and neglecting 
the NNLO correction from hard spectator scattering. 
Large perturbative logarithms in the hard-scattering 
kernels are identified and resummed using
soft-collinear effective theory.  We use our results to estimate  
the branching fractions for $B\to K^*\gamma$ and $B_s\to \phi\gamma$
decays at NNLO and compare them with the current experimental data.

\end{titlepage}

\newpage

\section{Introduction} 
%
Radiative $B\to V\gamma$ decays, where $V$ is a light 
vector meson,  are processes of particular interest 
in flavor physics which are already accessible 
at the $B$-meson factories at SLAC and KEK; 
current measurements~\cite{Coan:1999kh,Aubert:2004te,%
Nakao:2004th,Abe:2005rj,unknown:2006ag,Nakao-LP07,HFAG} 
yield the branching fractions presented 
in Table~\ref{tab:exp-data}. 
These decays provide independent constraints on the 
shape of the unitarity triangle, 
determining the side~$R_t$ of this triangle through 
the ratio of branching fractions for 
$B \to (\rho,\omega)\gamma$ and $B \to K^*\gamma$ decays.  
This information is complementary to the constraints 
on the ratio of  CKM matrix elements $|V_{td}/V_{ts}|$
obtained from the recent CDF measurement of the mass 
difference~$\Delta M_s$ in the $B_s - \bar B_s$ 
system~\cite{Abulencia:2006ze} and the already precise knowledge 
of the $B_d - \bar B_d$ mass difference $\Delta M_d$~\cite{HFAG}.
Moreover, measurements of the CP-asymmetries in 
$B \to (\rho,\omega)\gamma$ decays and the isospin-violating 
ratio of the charged and neutral 
$B \to \rho\gamma$ modes would determine the inner 
angle~$\alpha$ of the unitarity triangle. 

The calculation of the branching fractions for $B\to V\gamma$ 
decays requires the evaluation of the hadronic matrix elements 
of the operators in the effective weak Hamiltonian.
For $B\to V\gamma$ 
decays the weak Hamiltonian is~\cite{Buchalla:1996vs,Beneke:2000ry}: 
\begin{equation}
\mathcal{H}_{\rm eff} = 
\frac{G_F}{\sqrt 2} \sum_{p=u,c} \lambda_p^{(q)}  
\left [ C_1  Q_1^p + C_2  Q_2^p + 
\sum_{i=3}^8 C_i Q_i \right ] , 
\label{eq:heff}
\end{equation}
where $\lambda_p^{(q)} = V^*_{pq} V_{pb}$ 
(unitarity of the CKM matrix implies that
$\lambda_t^{(q)} = - (\lambda_u^{(q)} + \lambda_c^{(q)})$ 
and so contributions from diagrams with top-quark loops
are included implicitly). The relevant four-quark operators 
$Q_1$ and $Q_2$ are
\begin{eqnarray}
\label{eq:4-quark-operators}
 Q_1^p = (\bar q \, p)_{V-A} \, (\bar p \, b)_{V-A}, 
& \qquad &
 Q_2^p = (\bar q_i p_j)_{V-A} \,(\bar p_j b_i)_{V-A}, 
\end{eqnarray}
and the electromagnetic and chromomagnetic penguin 
operators $Q_7$ and $ Q_8$ are
\begin{eqnarray}
 Q_7 = -\frac{e \, \overline m_b(\mu)}{8\pi^2}\,
\bar q \, \sigma^{\mu\nu} \, [1 + \gamma_5] \, b 
 F_{\mu\nu} \, , 
& \quad &
 Q_8 = -\frac{g \, \overline m_b(\mu)}{8\pi^2}\,
\bar q \,\sigma^{\mu\nu} \, [1 + \gamma_5] \, T^a \, b 
 G^a_{\mu\nu} . 
\label{eq:penguin-operators} 
\end{eqnarray}
Here $q = d$ or~$s$, and the convention for the sign of the 
couplings corresponds to the covariant derivative
$iD_\mu = i\partial_\mu +  e Q_f A_\mu +  g T^a A_\mu^a$, 
with~$A_\mu$ and~$A_\mu^a$ representing the photon and gluon 
fields respectively, and $Q_e = -1$ etc.  The factor
$\overline m_b(\mu)$ is the $\MSbar$ mass of the $b$ quark.
The Wilson coefficients~$C_i$ have been 
known within the next-to-leading logarithmic approximation
(NLL) for over a decade (for a review, see
\cite{Hurth:2003vb}), and have
been recently calculated at next-to-next-to-leading logarithmic
order (NNLL) in a series of papers
\cite{Bobeth:1999mk,Misiak:2004ew,Gorbahn:2004my,Gorbahn:2005sa,Czakon:2006ss}.
In the present work we focus on the most phenomenologically relevant 
operators, which are $Q_1$,  $Q_7$, and $ Q_8$.  
The matrix elements of the QCD-penguin operators  $Q_3, \, \ldots Q_6$ 
first contribute at~${\cal O} (\alpha_s)$ and are 
multiplied by small Wilson coefficients in the weak 
Hamiltonian $\mathcal{H}_{\rm eff}$~(\ref{eq:heff}).  
The contribution from $Q_2$ starts at ${\cal O} (\alpha_s^2)$.

\begin{table}[t]
\caption{
Status of the $B$-meson radiative branching fractions
(in units of $10^{-6}$) from the BABAR, BELLE and CLEO collaborations and
their averages by HFAG~\cite{HFAG}. The entry for $B_s \to \phi \gamma$
is from the recent BELLE measurement~\cite{Nakao-LP07}.
}
\label{tab:exp-data}
\begin{center}
\begin{tabular}{l|ccc|c}
\hline
Mode & BABAR & BELLE & CLEO & HFAG
\\ \hline
$B^+ \to K^{*+} \gamma$ &
$38.7 \pm 2.8 \pm 2.6$ &
$42.5 \pm 3.1 \pm 2.4$ &
$37.6^{+8.9}_{-8.3} \pm 2.8$ &
$40.3 \pm 2.6$ \\
$B^0 \to K^{*0} \gamma$ &
$39.2 \pm 2.0 \pm 2.4$ &
$40.1 \pm 2.1 \pm 1.7$ &
$45.5^{+7.2}_{-6.8} \pm 3.4$ &
$40.1 \pm 2.0$ \\ \hline
$B^+ \to \rho^+ \gamma$ &
$1.10^{+0.37}_{-0.33} \pm 0.09$ &
$0.55^{+ 0.42 + 0.09}_{- 0.36 - 0.08}$ &
$< 13$ &
$0.88^{+0.28}_{-0.26}$ \\
$B^0 \to \rho^0 \gamma$ &
$0.79^{+0.22}_{-0.20} \pm 0.06$ &
$1.25^{+ 0.37 + 0.07}_{- 0.33 - 0.06}$ &
$< 17$ &
$0.93^{+0.19}_{-0.18}$ \\
$B^0 \to \omega \gamma$ &
$0.40^{+0.24}_{-0.20} \pm 0.05$ &
$0.56^{+ 0.34 + 0.05}_{- 0.27 - 0.10}$ &
$< 9.2$ &
$0.46^{+0.20}_{-0.17}$ \\ \hline
%
%
$B \to K^* \gamma$ &
$40.4 \pm 2.5$ &
$42.8 \pm 2.4$ &
$43.3 \pm 6.2$ &
$41.8 \pm 1.7$ \\
$B \to (\rho,\omega) \, \gamma$ &
$1.25 \pm 0.25 \pm 0.09$ &
$1.32^{+0.34 +0.10}_{-0.31 -0.09}$ &
$< 14$ &
$1.28^{+0.31}_{-0.29}$ \\ \hline
$B_s \to \phi \, \gamma$ & & $ 57^{+18 +12}_{-15 -17}$ & &
\\ \hline
\end{tabular}
\end{center}
\end{table}

It has been shown that in the heavy-quark limit 
a factorization framework (called QCD factorization) can be
applied to $B\to V\gamma$ decays~\cite{
Ali:2001ez,Ali:2004hn,Beneke:2001at,Kagan:2001zk,Bosch:2001gv, 
Bosch:2004nd,Beneke:2004dp} (see \cite{Ali:2006fa,Ball:2006eu}
for phenomenological updates to NLO, and ~\cite{Keum:2004is,Lu:2005yz}
for the alternative ``perturbative QCD'' approach). In particular,
the matrix element of a given operator in the effective weak
Hamiltonian  can be written in the form
\begin{equation} 
\left \langle V \gamma \left | Q_i 
\right | \bar B \right \rangle = 
F^{B \to V_\perp}  \, T_i^{\rm I} + 
\int d\omega \, du \,
\phi^B_+ (\omega) \, \phi^V_\perp (u) \, 
T^{\rm II}_i (\omega,u) \, .
\label{eq:ff} 
\end{equation}
The non-perturbative effects are contained 
in $F^{B \to V_\perp}$, the $B \to V$ transition 
form factor at $q^2 = 0$, and 
in $\phi^B_+$ and $\phi^V_\perp$, 
the leading-twist light-cone distribution amplitudes
(LCDAs) of the $B$- and $V$-mesons. The hard-scattering 
kernels~$T^{\rm I}_i$ and~$T^{\rm II}_i$ include 
only short-distance effects and are calculable in 
perturbation theory. Contributions to the kernel
$T^{\rm I}$ are closely related to the virtual corrections 
to the inclusive decay rate, and are referred to as vertex corrections.  
Those to the kernel $T^{\rm II}$ are
related to parton exchange with the light quark in the 
$B$-meson, a mechanism  commonly referred to as 
hard spectator scattering.  It is expected that the factorization 
formula is  valid up to corrections of ${\cal O}(\Lambda_{\rm QCD}/m_b)$. 

The derivation of the factorization formula from a two-step
matching procedure in soft-collinear effective theory
 (SCET)~\cite{Bauer:2000ew,Bauer:2000yr,Bauer:2001yt,Beneke:2002ph} 
has provided  additional insight into its structure.   
An advantage of the effective field-theory approach 
is that it allows for an unambiguous
separation of scales and an operator definition of 
each object in the factorization formula.
The technical details for $B\to V\gamma$ have been provided
in \cite{Chay:2003kb,Grinstein:2004uu,Becher:2005fg}.  In the SCET 
approach the factorization formula is written as 
\begin{equation}
\label{eq:SCETff}
\left \langle V \gamma \left |  Q_i 
\right | \bar B \right \rangle  = \Delta_i C^{A}\zeta_{V_\perp}  + 
\frac{\sqrt{m_B}F f_{V_\perp}}{4}\int d\omega \, du \,
\phi^B_+ (\omega) \, \phi^V_\perp (u) \, 
t^{\rm II}_{i} (\omega,u)\,,
\end{equation}
where $F$ and $f_{V_\perp}$ are meson decay constants.
The SCET form factor $\zeta_{V_\perp}$ is related to the QCD form 
factor through perturbative and power corrections
\cite{Beneke:2000wa,Beneke:2003pa,Lange:2003pk,Beneke:2004rc,
Beneke:2005gs,Hill:2004if,Becher:2004kk}.  
In SCET the perturbative hard-scattering kernels are the 
matching coefficients $\Delta_i C^A$ and 
$t_i^{\rm II}$.  They are known completely to next-to-leading order
(NLO) $({\cal O}(\alpha_s))$ in renormalization-group (RG) 
improved perturbation theory \cite{Becher:2005fg}. 
In this paper we make steps towards a complete analysis
at next-to-next-to-leading order (NNLO) by obtaining 
full results for the hard-scattering kernels
for the dipole operators $Q_7$ and $Q_8$, and partial results 
for $Q_1$, valid in the large-$\beta_0$ limit and neglecting
NNLO corrections from spectator scattering.

The hard-scattering kernels are found 
by matching certain partonic matrix elements in QCD with those in the
effective theory. For the vertex corrections 
the relevant matrix elements are $\langle s \gamma|Q_i|b\rangle$.
The loop corrections in the effective theory can be made to 
vanish by matching on-shell, so the main obstacle is 
the evaluation of the QCD matrix  elements.  However, 
these matrix elements are just the virtual corrections 
to the inclusive  $B \to X_s \gamma$ decay rate.  
Exact results to ${\cal O}(\alpha_s^2)$  were obtained for 
$Q_7$ in~\cite{Blokland:2005uk,Asatrian:2006ph} and for 
$Q_8$ in~\cite{greub_prep}.
For $Q_1$ the virtual corrections at  ${\cal O}(\alpha_s)$ were 
calculated in~\cite{Greub:1996tg,Buras:2001mq,Buras:2002tp}, 
but those at ${\cal O}(\alpha_s^2)$
are known only in the large-$\beta_0$ 
limit~\cite{Bieri:2003ue,Boughezal:2007ny}\footnote{these results are obtained by 
calculating the ${\cal O}(\alpha_s^2 n_f)$ terms and
then replacing $n_f\to -3\beta_0/2$, according 
to the hypothesis of ``naive non-abelianization'' \cite{Beneke:1994qe}.}. 
A calculation that goes beyond this
approximation by employing an interpolation in the charm quark mass $m_c$
was reported in ~\cite{Misiak:2006ab}, and has  been used in estimating
the NNLO branching fraction for the inclusive decay
 $B \to X_s \gamma$~\cite{Misiak:2006zs}. However, as the calculation
was not split into virtual and bremsstrahlung contributions, those
results cannot be used in the SCET matching calculation.
Therefore, while we obtain exact NNLO results $Q_7$ and $Q_8$,
for $Q_1$ we are restricted to the large-$\beta_0$ limit.
Our results provide an explicit check 
on factorization at NNLO.

Corrections from spectator scattering are included in the 
hard-scattering kernel $t^{\rm II}$ and first contribute to the 
branching fraction at NLO $({\cal O}(\alpha_s))$.  
A complication of spectator scattering is the 
presence of two widely separated perturbative 
scales $m_b^2\gg m_b \Lambda_{\rm QCD}$.
The SCET approach provides a systematic framework for 
separating contributions from these two scales.  
In SCET the hard-scattering kernel $t_i^{\rm II}$ for a given
operator is sub-factorized into the convolution of a hard-coefficient 
function with a universal jet function, in the form
\begin{equation}\label{eq:CJ}
t_i^{{ \rm II}}(u,\omega)=\int_0^1 d\tau \Delta_i C^{B1}(\tau) 
j_\perp(\tau,u,\omega)
\equiv  \Delta_i C^{B1}\star j_\perp.
\end{equation}
The hard coefficients $\Delta_i C^{B1}$ contain 
physics at the hard scale $m_b$, while 
the jet function $j_\perp$ contains physics at the hard-collinear 
scale $\sqrt{m_b \Lambda}$.  The hard coefficient is 
identified in a first step of matching ${\rm QCD} \to{\rm SCET}_{\rm I}$, 
and the jet function in a second step of matching 
${\rm SCET}_{\rm I}\to {\rm SCET}_{\rm II}$. Details have been 
worked out for  $B\to V\gamma$ in  
\cite{Chay:2003kb,Becher:2005fg}, for heavy-to-light form factors in 
\cite{Bauer:2002aj,Beneke:2003pa,Lange:2003pk,Beneke:2004rc,Beneke:2005gs, 
Hill:2004if, Becher:2004kk}, and for $B\to PP$ 
in \cite{Beneke:2005vv,Beneke:2006mk}.

The effective field-theory techniques are crucial for
providing a field-theoretical definition of the objects 
in (\ref{eq:SCETff}),  and for resumming large perturbative
logarithms of the ratio $m_b/\Lambda_{\rm QCD}$ in the $t_i^{\rm II}$.  
In the effective-theory approach resummation is carried out
by solving the renormalization-group equations for the matching
coefficients $\Delta_i C^{B1}$. Since these coefficients
enter the factorization formula in a convolution with the jet
function $j_\perp$, their anomalous dimension is a distribution 
in the variables $\tau$ and $u$. The evolution equations must be solved 
before performing the convolution with $j_\perp$.  Therefore,
resummation is not possible in the original QCD 
factorization formula (\ref{eq:ff}), where 
the hard-scattering kernels $T_i^{\rm II}$ are obtained
only after this convolution has been  carried out.

While the SCET formalism is indispensable for resummation,     
in the actual matching calculations one can also use the 
diagrammatic method of expanding by regions \cite{Beneke:1997zp} 
in order to separate hard from hard-collinear effects as in
(\ref{eq:CJ}).  This method was used to analyze loop
corrections to spectator scattering for the case of the $B\to \pi$ 
form factor in \cite{Kirilin:2005xz},  and for 
$B\to \pi\pi$ in \cite{Kivel:2006xc}. In
both cases the results were shown to be equivalent to those
obtained directly in SCET.  We use similar techniques here to compute  
the NNLO correction from the hard-scattering kernel $t_8^{\rm II}$.  
Our result for the one-loop correction at the hard-collinear 
scale agrees with (\ref{eq:CJ}), explicitly confirming
the universality of the jet function predicted by SCET.
Since the  NNLO corrections from $t_7^{\rm II}$ are known from the 
form factor analysis  \cite{Becher:2004kk,Beneke:2005gs}, the main
obstacle to a complete treatment of spectator scattering is the 
NNLO matching calculation for $Q_1$. 

The paper is organized as follows. In Section \ref{sec:HS-kernel}
we explain the SCET factorization framework and define the 
hard-scattering kernels. The SCET matching calculations
are carried out for the vertex corrections
in Section \ref{sec:Vertex} and for the hard-spectator
corrections in Section \ref{sec:spectator}. In Section  
\ref{sec:numerics} we describe the numerical analysis and estimate
the branching fractions for $B\to K^\star\gamma$ and $B_s\to\phi\gamma$
decays at NNLO, comparing our results with the current data and 
identifying the theoretical uncertainties.  
We conclude in Section \ref{sec:Conclusions}.
Results for the partonic matrix elements taken from calculations
for inclusive $B\to X_s\gamma$ decays are relegated to the Appendix,
along with some of the SCET matching
functions obtained in previous work and details of the 
renormalization-group analysis.

\section{Factorization and the  hard-scattering kernels}
\label{sec:HS-kernel} 

In this section we explain the closely related issues of 
factorization and extraction of the hard-scattering kernels.  
The objects of interest are the hadronic matrix elements
$$\left \langle V \gamma \left | Q_i 
\right | \bar B \right \rangle. $$
An analysis in \cite{Becher:2005fg} used a two-step matching
procedure in SCET to show that these
matrix elements can be written in the form (\ref{eq:SCETff}) to all
orders in perturbation theory and to leading order in
$1/m_b$. In this paper we work out a large set of effective-theory 
matching coefficients at NNLO in perturbation theory. 
These are obtained by replacing the hadronic states by partonic ones and 
calculating the matrix elements in perturbative QCD.  Showing that
the partonic rate can be brought into the form (\ref{eq:SCETff})
demonstrates factorization and provides expressions  
for the hard-scattering kernels.

To calculate the partonic matrix elements requires the evaluation of 
multi-scale Feynman integrals.  It is advantageous to 
perform these integrals using the method of regions \cite{Beneke:1997zp}.  
This not only provides a simple way to obtain results at leading order
in $1/m_b$, but also a factorization of momentum 
scales at the level of Feynman diagrams.   
In this method the loop integrations are split into a sum of different
regions, in which the loop momenta satisfy a fixed scaling.  This
allows for a Taylor expansion under the integral in each 
region, which is subsequently integrated over all space.  The integrals 
are performed in dimensional regularization, where scaleless integrals
are set to zero.  The sum of the results for all the regions 
recovers the full integral, expanded in $1/m_b$.  

A number of different momentum regions appear in the analysis, 
both perturbative and non-perturbative.
To identify these we first introduce two light-like vectors
$n_{\pm}$ satisfying $n_+n_-=2$.  We choose the outgoing
vector meson to travel along the $n_-$ direction, and define $n_+$
such that the velocity of the $b$ quark is given by
\begin{equation}
v^{\mu}= n_-^{\mu}\frac{n_+v}{2}+ n_+^{\mu}\frac{n_-v}{2}.
\end{equation}
This definition implies $v_\perp=0$, and we shall always work
in the reference frame where $n_-v=n_+v=1$. To perform the expansion 
in $1/m_b$, we define the parameter
$\Lambda^2=(p_B-m_b v)^2$ and the dimensionless parameter
$\lambda=\Lambda/m_b \ll 1$. The regions
are classified according to the scaling of their light-cone
components with the expansion parameter $\lambda$.
Denoting the light-cone components of a 
generic four-vector $p$ by $(n_+p,p_\perp,n_-p)$,  
the relevant momentum regions are \cite{Becher:2005fg}:  

%
\begin{center}
\begin{tabular}{ll}
\underline{ \bf Perturbative} & 
\\
&\\
hard  & \hspace{.5cm}  $m_b(1,1,1)$ 
\\ 
hard-collinear   & \hspace{.5cm}
$m_b(1,\sqrt{\lambda}, \lambda)$ \\
& \\
\underline{{\bf Non-perturbative}} & \hspace{.3cm}
\\ & \\
soft & \hspace{.5cm}
$m_b (\lambda,\lambda,\lambda)$\\ 
collinear  &\hspace{.5cm}
$m_b (1,\lambda,\lambda^2)$ \\
soft-collinear & \hspace{.5cm}
$m_b (\lambda,\lambda^{3/2},\lambda^2)$
\end{tabular}
\end{center}
The connection between the SCET analysis and perturbative QCD is provided
by the method of regions.  In the effective theory, 
contributions from the perturbative regions are encoded in 
Wilson coefficients of operators built
from fields representing the regions of lower virtuality. 
It is convenient to factorize the two perturbative
scales $m_b^2$ and $m_b \Lambda$ using a two-step matching procedure
${\rm QCD}\to {\rm SCET_I}\to {\rm SCET_{II}}$.

In the first matching step the hard scale $m_b^2$ is integrated
out by matching the operators $Q_i$ 
onto a set of operators in ${\rm SCET_I}$.  
The effective theory ${\rm SCET_I}$ involves fields for the
hard-collinear and non-perturbative modes, multiplied
by  Wilson coefficients related to the hard region.  For the case of 
$B\to V\gamma$, the matching takes the  
form \cite{Becher:2005fg}
\begin{equation}
 Q_i \to 
\Delta_i C^{A}  J^{A}
+ \Delta_i C^{B1}\star J^{B1}+ \Delta_i C^{B2}\star J^{B2}.
\end{equation}
The $\star$ denotes a convolution over momentum fractions, as 
in (\ref{eq:CJ}).  The momentum-space Wilson coefficients depend only 
on quantities at the hard scale $m_b^2$. 
The exact form of the operators $J^{(i)}$ along with the relevant
SCET conventions can be found in  \cite{Becher:2005fg}:
\begin{eqnarray}
\label{eq:SCETJs}
J^{A}&=&\left(\bar\xi W_{hc}\right)\epsslash_\perp(1-\gamma_5) h_v, \\
J^{B1}&=&\left(\bar\xi W_{hc}\right)\epsslash_\perp \calAslash_{hc_\perp} 
(1+\gamma_5)  h_v ,\\
J^{B2}&=& \left(\bar\xi W_{hc}\right)\calAslash_{hc_\perp}  \epsslash_\perp 
(1+\gamma_5)  h_v.
\end{eqnarray}
Here $\varepsilon_\perp$ is the polarization vector of the 
on-shell photon. The operators contain a hard-collinear
quark field $\xi$, a composite object ${\cal A}_{hc}$, which in
light-cone gauge is the hard-collinear gluon field, 
and $W_{hc}$, a Wilson line.  In SCET the $b$-quark field
is treated as in HQET. We have suppressed the arguments of 
the fields above, but must keep in mind that due to the non-locality of SCET
the objects $\left(\bar\xi W_{hc}\right)$ and $\calAslash_{hc_\perp}$ are 
evaluated at different points along the $n_+$ light-cone,
whereas $h_v$ is multipole expanded and evaluated at a point
on the $n_-$ light-cone (see, e.g., \cite{Beneke:2002ph}).
The $B$-type operators are actually power 
suppressed in ${\rm SCET_I}$, but contribute at the same order 
as the $A$-type operator upon the transition to  
${\rm SCET_{II}}$  \cite{Bauer:2002aj,Beneke:2003pa,Lange:2003pk}.

The matrix element of the operator 
$J^{A}$ is proportional to the SCET form factor $\zeta_{V_\perp}$.  
The Wilson coefficients $\Delta_i C^A$ multiplying this matrix
element can be extracted from calculations in the inclusive 
$B\to X_s \gamma$ decay.  Details are given in 
Section \ref{sec:Vertex}.  In contrast to the QCD form factor, 
the SCET form factor contains no piece which can be written 
in the form of a (convergent) convolution of a hard-scattering 
kernel with the meson LCDAs 
\cite{Beneke:2003pa, Lange:2003pk}\footnote{although see \cite{Manohar:2006nz}
for a renewed discussion of this point.}.   The relation between the 
QCD form factor and the SCET form factor is determined by 
the factorization formula~\cite{Beneke:2003pa,Lange:2003pk,Beneke:2000wa}
\begin{equation}
\label{eq:ffForm}
F^{B\to V_\perp}= C_{V_\perp}^{A} \zeta_{V_\perp} + 
\frac{\sqrt{m_B}F f_{V_\perp}}{4} \int d\omega \, du \,
\phi^B_+ (\omega) \, \phi^V_\perp (u) \, 
t^{\rm II}_{V_\perp} (\omega,u)\,.
\end{equation}
Since the matrix element of $Q_7$ is proportional to the form
factor, the coefficient functions $C_{V_\perp}^A$ and 
$t_{V_\perp}^{\rm II}$ at NNLO can be determined from the results 
for $Q_7$.  The exact relation is given in (\ref{eq:Form}) below.  

The operators $J^{(Bi)}$ can be further matched onto four-quark
operators in ${\rm SCET_{II}}$. 
For $B\to V \gamma$ decays, only the operator $J^{B1}$ is relevant.
The matrix element of the four-quark operator onto which it 
matches factorizes into a product of LCDAs for the $B$
and $V$ mesons.  The operator $J^{B2}$, on the other hand,
matches onto a four-quark operator whose renormalized matrix element
has no projection on the pseudoscalar $B$-meson LCDA.
In matching the operator $J^{B1}$ onto ${\rm SCET_{II}}$
the hard-collinear scale $m_b \Lambda$ is integrated out, 
and the associated Wilson coefficient is the jet function 
$j_\perp$.  
The final low-energy theory ${\rm SCET_{II}}$ contains only soft, collinear,
and soft-collinear fields.  Factorization means that
soft fields are restricted to the $B$-meson LCDA, and  
collinear ones to the $V$-meson LCDA. Since these two
pieces communicate only through soft-collinear interactions,
factorization amounts to showing that such contributions decouple from the 
hadronic matrix element of the ${\rm SCET_{II}}$ operator. 
This was done in \cite{Becher:2005fg}.  
Thus the matrix element of the  operator onto which 
$J^{B1}$ matches is exactly of the  form of the second 
piece of (\ref{eq:SCETff}), with $t_i^{\rm II}=\Delta_i C^{B1}\star j_\perp$.
This same jet function appears in the factorization 
formula (\ref{eq:ffForm}) for the form factor, 
where $t^{\rm II}_{V_\perp}=C^{B1}_{V_\perp}\star j_\perp$.
We can summarize this discussion by the following factorization
formula
\begin{eqnarray}
\label{eq:SCETffQCDff}
\left \langle V \gamma \left | Q_i
\right | \bar B \right \rangle
&=& \Delta_i C^A \zeta_{V_\perp}+\frac{\sqrt{m_B}F f_{V_\perp}}{4}
\left(\Delta C^{B1}\star j_\perp\right)
\star  \phi^V_\perp \star\phi^B_+
\\
&=& \frac{\Delta_i C^{A}}{C_{V_\perp}^A} F^{B\to V_\perp}
+\frac{\sqrt{m_B}F f_{V_\perp}}{4}
\left[\left(\Delta_i C^{B1}-\frac{\Delta_i C^A}{C_{V_\perp}^A} 
 C^{B1}_{V_\perp}\right)\star j_\perp\right]
\star  \phi^V_\perp \star\phi^B_+ . \nonumber
\end{eqnarray}
This formula relates the hard-scattering kernels $\Delta_i C^A$
and $t_i^{\rm II}$ in (\ref{eq:SCETff}) to the  
Wilson coefficients from  the two-step matching procedure in SCET,
and provides a connection with the original formulation
(\ref{eq:ff}). For instance,  using that 
$\Delta_7 C^i\sim C^i_{V_\perp}$,  one can verify that 
$Q_7$ contributes to both terms in the SCET formulation,
but only to the vertex term in the original formulation. 

A main result of our paper is an expression for 
the ${\cal O}(\alpha_s^2)$ correction to the
hard coefficient $\Delta_8 C^{B1}$. We obtain it 
with a straightforward diagrammatic analysis 
using the method of regions, without the explicit formulation of
SCET or the use of its Feynman rules. 
Since contributions from $J^{B1}$ can be uniquely identified
by the Dirac structure of the four-quark operator onto which it matches, the 
sub-factorization of the hard-scattering kernel into a convolution
of a jet and hard function can be performed by separating
out the contributions of the hard and hard-collinear regions 
multiplying this structure.  Details 
are given in Section \ref{sec:spectator}.

\section{Vertex corrections}
\label{sec:Vertex}
We begin with the vertex corrections, extracting the contributions of 
the operators $Q_1,Q_7$, and $Q_8$ to the SCET Wilson 
coefficient $C^A$ at NNLO (${\cal O}(\alpha_s^2))$.
To do so we calculate the partonic matrix 
elements $$\langle Q_i \rangle \equiv
\langle q(p)\gamma(q)|Q_{i}|b(p_b)\rangle$$
to this same order in both SCET and QCD.  
This matrix element is chosen because it contains no external gluons 
and so matches directly onto the operator $J^{A}$ in (\ref{eq:SCETJs}).
The calculation is performed with on-shell external 
quark states and  both UV and IR  divergences are regularized dimensionally.
In that case the matching calculation is simple, 
because the loop corrections in SCET are scaleless and vanish. The 
matrix element of $J^A$ is just the tree expression plus counterterms
from wave-function and current renormalization.
The QCD matrix elements can be read off from the virtual corrections 
to the inclusive decay $B\to X_s \gamma$.
Using that the on-shell wave-function 
renormalization factors in the effective theory
are unity, and replacing the bare SCET current by its 
renormalized one, we have 
\begin{equation}
\langle  Q_i \rangle= 
D_i\langle Q_{7,{\rm tree} }\rangle = \Delta_i C^{A}
Z_J\langle J^{A}_{\rm tree} \rangle. 
\end{equation}
Here the $D_i$ are the scalar amplitudes in QCD, the  $\Delta_i C^{A}$
are the contributions of a given operator to the SCET matching coefficient,
and $Z_J$ is the renormalization factor of the SCET current
operator $J_A$.  Each of these
quantities is determined as a series in $\alpha_s$.  For the 
operators $Q_{7,8}$ we can obtain complete results at NNLO, 
while for $Q_1$ we can only provide an estimate using 
the large-$\beta_0$ limit.

We first consider tree level, where only $Q_7$ contributes.   
For on-shell matching the spinors in QCD and SCET are 
equal to one another and we  find
\begin{equation}
\Delta_7 C^{A(0)}= -\frac{e\,{\overline m}_b \,2E_\gamma}{4\pi^2} ,
\end{equation}
where the photon energy is $2E_\gamma=m_B(1-m_V^2/m_B^2)\approx m_b$ in 
the heavy-quark limit.
At higher orders the matching coefficients can be read off 
from the  functions $D_i$  according to the relation
\begin{equation}
  \Delta_i C^{A}(m_b,\mu) = \Delta_7 C^{A(0)}
\lim_{\epsilon\to 0}\,Z_J^{-1}(\epsilon,m_b,\mu)\,
   D_{i}(\epsilon,m_b,\mu) \,.
\end{equation}
The SCET current renormalization factor 
$Z_J$ is determined by requiring that the Wilson coefficient
be free of IR poles.  

Before giving results for the higher-order corrections, we pause to
explain a subtlety in the matching which 
first appears at two loops.  The on-shell matrix elements of the 
QCD operators $Q_i$ are calculated in $\MSbar$ renormalization in
the five-flavor theory,  $n_f=n_l+n_h$ with $n_h=1$ for the $b$ 
quark.  However, in SCET $b$-quark loops are absent and the 
matrix elements are calculated as an expansion in the four-flavor theory. 
In order to perform a correct matching, it is necessary to express
the UV renormalized results in the five-flavor theory in terms
of the four-flavor parameters of SCET.
A similar problem arises when integrating out the top quark
to match the Standard Model onto the effective weak
Hamiltonian. The solution is to  renormalize the
coupling constant in the $n_f=n_h+n_l$ flavor
theory according to $\alpha_s^{\rm bare}=Z_{\alpha}^{n_h+n_l}\alpha_s$, 
with~(see e.g. \cite{Misiak:2004ew, Steinhauser:2002rq})
\begin{equation}\label{eq:Znhnl}
Z_{\alpha}^{n_h+n_l}=1-\frac{\alpha_s}{4\pi \epsilon}
\left[\frac{11}{3}C_A-\frac{2}{3}n_f+ \frac{2}{3}n_h(1-N_\epsilon)\right].
\end{equation}
The function $N_\epsilon$ is fixed such that 
$\alpha_s$ is the $\overline{\rm MS}$-renormalized coupling
in the {\it four} flavor theory. Its value is
\begin{equation}
\label{eq:Znfnl}
N(\epsilon)=e^{\gamma\, \eps}
\left(\frac{ \mu^2}{m_b^2}\right)^\epsilon \Gamma(1+\epsilon)\,.
\end{equation} 
Results for the scalar amplitudes $D_i$ 
in this renormalization scheme can be obtained from
the $\MSbar$ results given in the Appendix by making the replacement 
\begin{equation}
\label{eq:decoupling}
\alpha_s \to 
\alpha_s\left(1+
\frac{ \alpha_s}{4\pi} \frac{4}{3} n_h\left[L+
\epsilon\left(L^2 +\frac{\pi^2}{24}\right)
+\epsilon^2\left( \frac{2L^3}{3}+\frac{\pi^2}{12} L- 
\frac{\zeta_3}{6}\right)\right]\right)
+\dots \, ,
\end{equation}  
where $L=\ln\mu/m_b$.
Note that this is just the standard decoupling relation when evaluated in
four dimensions. 

We now give results for the Wilson coefficients, which we write 
in the form
\begin{equation}
\Delta_i C^{A}= \Delta_7 C^{A(0)}\left[\delta_{i7} + 
\frac{\alpha_s(\mu)}{4\pi} \Delta_i C^{A(1)}+
\left( \frac{\alpha_s(\mu)}{4\pi} \right)^2 \Delta_i C^{A(2)} \right]\,.
\end{equation}
We begin with $Q_7$. Results can be given analytically, but since
those for $Q_8$ are only known numerically we treat $Q_7$ the same.
Using the scalar functions $D_7$ given in the Appendix we find
\begin{eqnarray}
\label{deltac7a}
\Delta_7 C^{A(1)}&=&C_F\left[-2 L^2-5L-2L_{\rm QCD} -6.8225\right], \nonumber \\
\Delta_7 C^{A(2)} &=& C_F^2\left(2L^4 
+ 14 L^3 +38.1449 L^2+56.14711 L +7.8159\right) \nonumber \\
&&+ C_F C_A\left(-4.8889 L^3 -33.9758 L^2- 92.3415 L - 83.8866\right)
\nonumber \\
&& +C_F n_l \left(0.8889 L^3+ 6.8889 L^2 + 19.9050 L + 23.8254\right)
\nonumber \\
&&+ C_F n_h \left(-1.3333 L^2+2.8889 L - 0.810288 \right),
\end{eqnarray}
where one is to use $n_l=4$ and $n_h=1$ 
in the above equation.   In the one-loop result 
we have distinguished the logarithms 
$L_{\rm QCD}=\ln \mu_{\rm QCD}/m_b$ and $L=\ln\mu/m_b$. 
The $\mu_{\rm QCD}$ dependence cancels against 
the scale dependence in the effective weak
Hamiltonian, whereas the $\mu$ dependence cancels against
the scale dependence of the SCET soft function $\zeta_{V_\perp}$ and
the running coupling constant.
At one loop it is straightforward to separate the logarithms by
identifying the UV and IR poles in the individual Feynman
diagrams.  At two loops the distinction can be made by using the 
renormalization-group equation (\ref{eq:ADdef}) below.  
We give explicit results for the case where $L$ is distinguished
from $L_{\rm QCD}$ in the Appendix, but in this 
section we quote the NNLO results only for $L=L_{\rm QCD}$.

We can use our results to determine the anomalous dimension 
of the operator $J^{A}$ up to two loops. 
The anomalous dimension is obtained from the coefficient 
$Z_J^{(1)}$ of the $1/\epsilon$ pole term in the current
renormalization factor and has the form
\begin{equation}
\label{eq:JAD}
\gamma^A=  2\alpha_s\,\frac{\partial}{\partial\alpha_s}\,Z_J^{(1)}(m_b,\mu)=
  - \Gamma_{\rm cusp}(\alpha_s)\,\ln\frac{\mu}{m_b} + \gamma^J(\alpha_s)
  \, ,
\end{equation}
where $\Gamma_{\rm cusp}$ is the cusp anomalous dimension appearing
in the renormalization-group theory of Wilson lines 
\cite{Korchemskaya:1992je} (it has recently been calculated to 
three loops \cite{Moch:2004pa}; the result is listed in the appendix).
The result for the renormalization factor to two loops is 
\begin{eqnarray}
&& Z_J = 1 +\frac{C_F\alpha_s}{4\pi}\bigg[-\frac{1}{\eps^2}
-\frac{5 }{2\eps}-\frac{2L}{\eps}\bigg] \nonumber \\
&& +C_F \left(\frac{\alpha_s}{4\pi}\right)^2\bigg[
-\frac{0.5 C_F}{\eps^4} +\frac{1}{\eps^3}
\bigg(-2.5 C_F + 2.75  C_A -0.5  n_l -2 C_F L\bigg) \nonumber \\
&&+\frac{1 }{\eps^2}
\bigg(-3.125 C_F + 3.5447 C_A - 0.5556 n_l
-2 C_F L^2 +
(-5 C_F +3.6667 C_A-0.6667 n_l) L \bigg) \nonumber \\
&& +\frac{1}{\eps}
\bigg(-2.6525 C_F -3.4386 C_A +1.9799 n_l +
(-4.1546 C_A + 1.1111 n_l)L\bigg)\bigg] \, ,
\end{eqnarray}
from which we find
\begin{eqnarray}
\label{eq:TwoLoopJAD}
\gamma^A&=&\frac{C_F \alpha_s}{4\pi}\left(-4 L - 5\right)
 \\&&
+C_F \left(\frac{ \alpha_s}{4\pi}\right)^2 
\left((-16.6183 C_A+4.444 n_l)L-10.6102 C_F
-13.7545 C_A +7.9195 n_l\right)\nonumber .
\end{eqnarray}
This is consistent with (\ref{eq:JAD}) and defines $\gamma^J$. 
The one-loop result was first  obtained in \cite{Bauer:2000yr}.
We note that in this case the $n_h$ dependence in the renormalization
factor $Z_J$ drops out after using (\ref{eq:decoupling}).  This must be the case, 
since in the effective-theory current the $b$ quark
is integrated out and so its anomalous dimension cannot depend
on $n_h$.  Our result for the anomalous dimension, along 
with the relation
\begin{equation}
\label{eq:ADdef}
\mu \frac{d}{d\mu} \Delta_i C^{A}= \gamma^A \Delta_i C^A \,,
\end{equation}
allows us to perform the separation of UV and SCET logs in the Wilson
coefficients given in the Appendix.

The same SCET current also appears in the study
of the inclusive $B\to X_s\gamma$ decay spectrum with a cut
on the photon energy \cite{Neubert:2004dd}.  A result 
equivalent to our two-loop matching coefficient 
$\Delta_7 C^{A(2)}$ with $\mu=m_b$ was recently obtained in
\cite{Becher:2006pu}.  Translating
our expression into the two-loop result for $h(1)$ given in 
\cite{Becher:2006pu}, we find numerical  agreement.  
The dependence on $n_h$ not taken into account in that work is  
negligible numerically. We can also check the two-loop anomalous
dimension by using RG-invariance of the inclusive decay rate
along with the anomalous dimensions of the jet and soft functions 
calculated in  \cite{Becher:2005pd, Becher:2006qw}. Here again
the results agree.

We repeat the calculation for $Q_8$. In this case the one-loop
result is IR finite. The two-loop matching equation also
becomes IR finite after the results are expressed in terms
of the renormalized current calculated above.
This is a check on the effective-theory
construction, according to  which the IR poles in the QCD amplitudes for 
each operator in the weak Hamiltonian are absorbed by the same SCET current.  
For the coefficient functions we find  
\begin{eqnarray}
\Delta_8 C^{A(1)}&=& C_F \left[2.6667 L_{\rm QCD} + 1.4734 + 2.0944 i\right], 
\nonumber \\
\Delta_8 C^{A(2)}&=&-C_F^2
\big[5.3333 L^3+ 32.2802 L^2 +50.9612 L+ 1.8875  \nonumber \\
&&+ i (4.1888 L^2 + 31.4159 L +29.8299)\big]
\nonumber \\
&&+  C_F C_A \big[15.1111 L^2 +31.6617 L+ 2.3833
   + i (23.7365 L +  28.0745)\big]\nonumber \\
&&- C_F n_l \big[1.7778 L^2 + 4.0386 L +1.7170 
+ i( 2.7925 L +4.4215)\big] \nonumber \\
&&+ C_F n_h \big[1.7778 L^2 -2.0741 L +0.8829 \big]. 
\end{eqnarray}

 Finally, we consider the four-quark operators $Q_1$ and
$Q_2$.  At NLO the contribution from $Q_1$ can be obtained
as an expansion in $m_c^2/m_b^2$, whereas that from $Q_2$ vanishes.
To extract the NNLO results for these operators would require
the QCD amplitudes $D_1$ and $D_2$ to this same order, which involves
the calculation of a large set of three-loop graphs.
These corrections are known exactly only in the large-$\beta_0$ limit, in 
an expansion in $z=m_c^2/m_b^2$\cite{Bieri:2003ue}.  Within
this approximation the result for $Q_2$ vanishes, and that for
$Q_1$ can be written as 
\begin{eqnarray}
\label{eq:c1exact}
&&\Delta_1 C^{A(1)}= \frac{m_b}{{\overline m}_b}C_F 
\left[-3.8519 L_{\rm QCD}+ r^{(1)}(z)\right],  \\
&&\Delta_1 C^{A(2)}= - \frac{3\beta_0}{2}\frac{m_b}{{\overline m}_b}C_F
\left[2.4691 L^2 + l^{(2)} (z) L + r^{(2)}(z) \right]\, , \nonumber
 \end{eqnarray}
where we have replaced $n_f\to -3\beta_0/2$ as 
appropriate in the large-$\beta_0$ limit. Within this limit
it is also consistent to set the ratio  $m_b/{\overline m}_b$
to unity, as we shall do in the numerical analysis of
Section \ref{sec:numerics}.
Since in the large-$\beta_0$ limit the amplitude is IR finite,
we can read off the functions $r^{(i)}$ and $l^{(2)}$ directly
from the results for inclusive $B\to X_s\gamma$ decay. Converting
to our notation we have
\begin{eqnarray}
r^{(1)}=\frac{r_2}{C_F}, \qquad
r^{(2)}=\frac{r_2^{(2)}}{C_F}\qquad
l^{(2)}&=&-\frac{l_2^{(2)}}{C_F} \, ,
\end{eqnarray}
where $r_2$ is defined in eq. (2.35) of \cite{Greub:1996tg},
and $r_2^{(2)}, l_2^{(2)}$ in eq. (22) of \cite{Bieri:2003ue}.
As an example, for  $m_c/m_b=1.2/4.8$ we have
\begin{eqnarray}
\label{eq:c1Annlo}
&&\Delta_1 C^{A(1)}= \frac{m_b}{{\overline m}_b}C_F 
\left[-3.8519 L_{\rm QCD} -3.4529 - 0.5138 i\right],  \\
&&\Delta_1 C^{A(2)}= - \frac{3\beta_0}{2}\frac{m_b}{{\overline m}_b}C_F
\left[2.4691 L^2 + 4.9083 L +
5.1203 +i (0.9953 L + 1.6014)\right]\, . \nonumber
 \end{eqnarray}

There are two major uncertainties associated  
the large-$\beta_0$ limit.  The first is that there is no 
way to quantify the size of the terms in 
$\Delta_1 C^{A(2)}$ not captured within this
limit. The second is that  the higher-order calculation
does not resolve the perturbative ambiguities in 
the ratios of quark masses $m_b/{\overline m}_b$ and  $m_c/m_b$
in the lower-order coefficient $\Delta_1 C^{A(1)}$:  
the difference between mass renormalization schemes
in these ratios is a correction proportional to $C_F \alpha_s$
and set to zero in the large-$\beta_0$ limit.
We discuss these uncertainties in more detail in
the numerical analysis of Section \ref{sec:numerics}.

In Section \ref{sec:numerics} we will be interested in
the dependence of the branching fractions on the choice 
of renormalization scales. Both the matching 
coefficients $\Delta_i C^{A}$ and the SCET soft function
$\zeta_{V_\perp}$ depend on the SCET factorization 
scale $\mu$.  It is convenient to use
the renormalization group to determine  the 
coefficients $\Delta_i C^{A}$ at an 
arbitrary scale $\mu$, given their value at a 
matching scale $\mu_h\sim \mu_{\rm QCD}\sim m_b$.
This allows us to fix $\mu=m_b$ and determine
the soft function $\zeta_{V_\perp}$ only at this single
scale. We can then study the dependence of the branching fractions
under variations in $\mu_h$ and $\mu_{\rm QCD}$, under
which it is formally invariant.  
The relevant RG formalism was worked out in \cite{Hill:2004if}. 
The expression we need is 
\begin{equation}
\label{eq:CAev}
\Delta_i C^{A}(m_b,\mu_h,\mu)= 
\left(\frac{m_b}{\mu_h}\right)^{a(\mu_h,\mu)} 
{\rm exp}{[S(\mu_h,\mu)+a_J(\mu_h,\mu)]}
\Delta_i C^{A}(m_b,\mu_{\rm QCD}=\mu_h,\mu_h)\,.
\end{equation}
In the above equation we have correlated the scales  
$\mu_{\rm QCD}=\mu_h$ for simplicity, although we can 
keep them separate using the results in the Appendix.
With this choice the dependence on $\mu_h=\mu_{\rm QCD}$ 
on the left-hand side cancels
against the dependence in the effective weak Hamiltonian, so that 
the branching fractions are invariant 
under variations of the matching scale $\mu_h$. 
The RG exponents $S$ and $a$, and $a_J$ are given by
\begin{eqnarray}
\label{eq:CuspExps}
S(\mu_1,\mu_2)&=& -\int_{\alpha_s(\mu_1)}^{\alpha_s(\mu_2)}
\frac {d \alpha}{\beta(\alpha)}\Gamma_{\rm cusp}(\alpha)
\int_{\alpha_s(\mu_1)}^\alpha 
\frac {d \alpha^\prime}{\beta(\alpha^\prime)},\\
a(\mu_1,\mu_2)&=&\int_{\alpha_s(\mu_1)}^{\alpha_s(\mu_2)}
\frac {d \alpha}{\beta(\alpha)}\Gamma_{\rm cusp}(\alpha), \\
 a_J(\mu_1,\mu_2)&=&\int_{\alpha_s(\mu_1)}^{\alpha_s(\mu_2)}
\frac {d \alpha}{\beta(\alpha)}\gamma_{J}(\alpha).
\end{eqnarray}
These exact solutions are  evaluated by expanding the 
anomalous dimensions and the QCD $\beta$-function 
as perturbative series in the strong coupling.  
We can do this to two-loop order for 
$a$ and $a_J$, and to three-loop order for $S$.
The expansions to this  order are listed in the Appendix.  

\section{Hard spectator scattering}
\label{sec:spectator}
In this section we consider the spectator scattering mechanism
and the calculation of $t_i^{\rm II}$ $(i=1,7,8)$. 
The leading corrections from spectator scattering contribute 
to the branching fractions
at NLO (${\cal O}(\alpha_s)$) and are known completely.  
The NNLO corrections from $Q_7$ are also
known \cite{Becher:2005fg}, since they can be taken  
from the heavy-to-light form factor analysis in
\cite{Becher:2004kk,Beneke:2004rc,Beneke:2005gs}.  
In this section we calculate the NNLO corrections from  $Q_8$.
We find agreement with a certain set of logarithmic corrections 
obtained in \cite{Descotes-Genon:2004hd}, and 
verify the important SCET result that contributions at the hard-collinear
scale for each operator in the effective weak Hamiltonian are taken 
into account by a universal jet function.   To complete the NNLO 
matching calculation for spectator scattering would require 
results for $Q_1$ and $Q_2$.  
This is a rather difficult calculation involving the evaluation 
of two-loop graphs depending on the ratio $m_c/m_b$.

Before presenting our results for $Q_8$, we first 
review the results for $Q_7$ as derived in \cite{Becher:2005fg}.  
This will fix some notation and  clarify the sub-factorization
of $t^{\rm II}_i$ into a convolution of hard and jet functions.
The calculation makes use of the two-step matching procedure 
outlined in Section \ref{sec:HS-kernel} to 
integrate out the perturbative scales  
$m_b^2\gg m_b\Lambda_{\rm QCD}$.  
At tree level and to leading order in the HQET expansion 
the result is 
\begin{equation}
t_7^{{\rm II}(0)}(u,\omega)= \int_0^1 d\tau \,\Delta_7 C^{B1(0)}(\tau)
j_\perp^{(0)}(\tau,u,\omega),
\end{equation}
where 
\begin{equation}\label{eq:leadingC7}
\Delta_7 C^{B1(0)}(\tau)=\frac{e \overline m_b }{4 \pi^2};
\qquad 
j_\perp^{(0)}(\tau, u,\omega) =-\frac{4\pi C_F\alpha_s}{N_c}
 \frac{1}{m_b \omega \bar u}\delta(\tau-u) \, .
\end{equation}
The one-loop correction to the hard-scattering kernel 
breaks into a sum of corrections to the hard coefficient
and the jet function according to
\begin{equation}
\label{eq:SplitSpect}
t_7^{{\rm II}(1)}= \Delta_7 C^{B1 (1)}\star j_\perp^{(0)} + 
\Delta_7 C^{B1 (0)}\star j_\perp^{(1)},
\end{equation}
where the superscripts denote the $(n)$-loop correction to 
each function and the $\star$ denotes a convolution over 
the variable $\tau$.  Explicit results for each term can 
be deduced from the form-factor 
analysis in \cite{Beneke:2004rc,Beneke:2005gs,Becher:2004kk}
and are listed in the Appendix.
Note that while the hard coefficient 
function $\Delta_7 C^{B1}$ is particular to the operator $Q_7$, 
the jet function $j_\perp$ is not.  It is determined by the matching step 
${\rm SCET}_{I}\to {\rm SCET}_{\rm II}$, which contains 
no information about the structure of the operators in the effective
weak Hamiltonian at the scale $m_b$.  In the SCET description
of spectator scattering, therefore, the non-trivial task 
is to determine the corrections at the hard scale $m_b$,
contained in the Wilson coefficients  $\Delta_i C^{B1}$.
The contributions at the hard-collinear scale
$m_b \Lambda$ can be obtained by performing the convolution
in the second term of (\ref{eq:SplitSpect}).  

In what follows we obtain an expression for $t_8^{{\rm II}(1)}$ in the
form (\ref{eq:SplitSpect}),  derived in the following way.   
We first calculate the hard-scattering kernel directly in 
QCD factorization,  but separate the contributions from the 
hard and hard-collinear scales using the method of regions.
We then show that the one-loop contribution
from the hard-collinear region is
exactly $\Delta_8 C^{B1(0)}\star j_\perp^{(1)}$.
Since both the coefficient function 
$\Delta_8 C^{B1(0)}\sim \bar \tau/\tau$ (with $\bar \tau \equiv 1-\tau$)
and the jet function $j_\perp^{(1)}$ are non-trivial functions 
of $\tau$, this provides a consistency check between the 
QCD factorization and the SCET formalism, and also a check
on our loop calculations.  The remaining contribution 
is from the hard region and is identified with   
$\Delta_8 C^{B1(1)}\star j_\perp^{(0)}$.  Since $j_\perp^{(0)}$
is a delta function in the variable $\tau$, this result
is sufficient to recover the coefficient function
 $\Delta_8 C^{B1(1)}$.  As mentioned in the Introduction,  
it is this $\tau$-dependent function which is needed to 
obtain the resummed hard-scattering kernel used in the
numerical analysis in Section \ref{sec:numerics}.

\subsection{$Q_8$ at tree level}
\label{sec:tree}

We start by reviewing the tree-level calculation. The strategy
is to  evaluate the partonic matrix element 
${\cal A}_8=\langle q (p_1) \,  \bar q^\prime (p_2)  \gamma(q) 
\, | \,  Q_8 \, | \, \bar q^\prime (k) \, b (p_b) \rangle$
and show that it can be written in the form (\ref{eq:SCETff}).
The hard-scattering kernel is independent of the exact choice
of the partonic momenta. We shall work with on-shell quarks in the 
initial and final states, and furthermore work in the 
reference frame where the perpendicular components 
of the external parton momenta vanish.
In this frame, the momenta can be chosen as $p_1=\Unew p,$
$p_2=\bar u p$, $k=\omega n_+/2$, $p_b=m_b v=p_B-k$, and 
$q=E_\gamma n_+$, with $\bar u\equiv 1-u$.  
At leading order in $1/m_b$ we can
write the vector-meson momentum as $p\approx m_b n_-/2$ and
the photon energy as $E_\gamma\approx m_b/2$. The photon's
polarization vector lies in the transverse 
plane and is denoted by $\epsilon_\perp$. 
The power counting is such that $\omega/m_b\sim \lambda \ll 1$.

\begin{figure}
\begin{center}
\includegraphics[width=.4\textwidth]{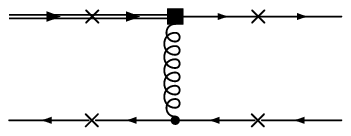}
\end{center}
\vspace{-.5cm}
\caption{The lowest-order diagram for spectator scattering
with $Q_8$.  The double-line represents the incoming
$b$ quark and the solid box  an insertion of $Q_8$.  The photon
can be attached to any of the four crosses.  Only photon
emissions from the light quark emerging from the $Q_8$ 
insertion contributes at leading power in $1/m_b$.}
\label{fig:tree8}
\end{figure}

The four Feynman diagrams which can contribute at tree-level 
are represented in Figure~\ref{fig:tree8}.  The photon can 
be emitted from any of the four crosses.   
At leading order in $\lambda$ only the diagram where the 
photon is attached to the light quark produced at the
flavor-changing weak current contributes. Emissions from the other quark lines
are either power suppressed or have no projection on the meson LCDAs.  
For the tree-level scattering amplitude one finds
\begin{eqnarray}
{\cal A}^{(0)}_8 & = & 
\frac{\overline m_b}{m_b}
\frac{e Q_d \alpha_s}{\pi} \, 
\frac{\bar u}{\Unew} \frac{1}{ \bar u \, \omega} \, 
\Big[ \bar u( \Unew p) \left\{ 
{\epsilon\llap{/}}_\perp \gamma_{\nu_\perp} \, ( 1 + \gamma_5 )  \right\}T^a u(p_b) \Big] \, 
\Big[ \bar v (k)  \gamma^\nu_{\perp}  
T^a v (\bar u p) \Big] \nonumber \\
& = & \frac{\overline m_b}{m_b}
\frac{C_F}{N_c} \frac{e Q_d \alpha_s}{\pi} \, \frac{\bar u}{\Unew} 
\frac{1}{ \bar u \, \omega} \,\left[{\epsilon\llap{/}}_\perp \gamma_{\nu_\perp} 
\otimes  \gamma^\nu_{\perp}\right]\equiv
A^{(0)} \,\left[{\epsilon\llap{/}}_\perp \gamma_{\nu_\perp} 
\otimes  \gamma^\nu_{\perp}\right],
\label{eq:A8} 
\end{eqnarray} 
where~$u$ and~$v$ represent the free-particle spinor 
wave-functions and $Q_d = -1/3$ denotes the charge of a 
down-type quark. We have distinguished the $\MSbar$ mass 
$\overline m_b(\mu)$ from the pole mass $m_b$, anticipating the one-loop 
calculation in the next section.  To obtain the second line 
we already performed the color trace, so that the 
notation $\Gamma_1 \otimes \Gamma_2$  
is to be understood as the Dirac structure between quark spinors.
In the second line we defined the tree-level partonic amplitude $A^{(0)}$.  
At tree level there is only one Dirac structure, 
related to the matching of the ${\rm SCET_I}$ operator $J^{B1}$. 

To proceed further we need definitions for the LCDAs in  
the low-energy theory ${\rm SCET}_{\rm II}$. 
The light-cone projection operator 
$\Phi^H_{\alpha\beta} (\tilde k)$ onto 
a heavy state~$H$ containing the~$b$-quark is given by 
\begin{equation} 
\Phi^H_{\alpha\beta} (\tilde \omega) = 
\int\, \frac{dt}{2\pi} \,  e^{i t \tilde \omega } 
\left\langle 0 \left |
\bar q_{s\beta} (tn_-) \left [ tn_-,0 \right ] h_{v\alpha} (0)
\right | H(v) \right\rangle \,,
\label{eq:phihdef}
\end{equation}
where~$q_s$ and~$h_v$ are soft and heavy quark fields in HQET; 
$\alpha$ and~$\beta$ are spinor labels. The quantity $[tn_-,0]$  
denotes a path-ordered exponential along the light cone.  
Similarly, the light-cone projection operator 
$\Phi^V_{\gamma\delta} (u)$ onto a light
meson state~$L$ is defined by
\begin{equation}
\Phi^L_{\gamma\delta} (u) = n_+ p 
\int\frac{ds}{2\pi} \, 
 e^{-i s \Unew n_+p  } 
\left\langle L (p) \left | 
\bar \xi_\delta (sn_+) \, [sn_+, 0] \, \xi_\gamma (0) 
\right | 0 \right\rangle \, ,
\label{eq:phimdef} 
\end{equation} 
where the $\xi$ are collinear quark fields in SCET.
The hadronic matrix elements of these light-cone projection
operators, contracted with certain Dirac structures,
are the LCDAs of the $B$ and $V$ mesons. The exact definitions of the 
distribution amplitudes needed in the analysis are
\begin{eqnarray}
\langle 0 |
\bar q_{s} (tn_-)  [ tn_-,0]\frac{\nslash_-}{2} h_{v} (0) | B(v) \rangle &=&
-\frac{i F(\mu)}{2}\sqrt{m_B} \,{\rm tr}
\big[\frac{\nslash_-}{2}\frac{1+\vslash}{2}\gamma_5\big]
\int_0^\infty \, d\omega e^{-i\omega t}\phi_+^B(\omega,\mu) \nonumber \\
\langle V (p)  | 
\bar \xi(sn_+) \, [sn_+, 0] \, \gamma_\perp^\mu \frac{\nslash_+}{2} \xi (0) 
 | 0 \rangle & = & \frac {i f_{V_\perp}(\mu)}{4} n_+ p \,
{\rm tr}\big[\etaslash_\perp \gamma_\perp^\mu \frac{\nslash_+\nslash_-}{4}\big]
\int_0^1 du \, e^{i s \Unew n_+p  } \phi_{\perp}^V(u,\mu), \nonumber \\
&&
\label{eq:HadLCDAs}
\end{eqnarray}
where $\eta_\perp$ is the polarization vector of the $V$-meson.

To extract the hard-scattering kernels we need only the 
partonic matrix elements. We write these as
a product of scalar distribution functions multiplied by appropriate
Dirac spinors. For on-shell matching at leading order in 
$1/m_b$ the QCD spinors are equal to the effective theory
ones.  At lowest order we have
\begin{eqnarray} 
\Phi^{b \bar q^\prime (0)}_{\alpha\beta} (\omega^\prime) 
& = &    \phi^{b \bar q^\prime (0)} \, 
\bar v_\beta (k) \, u_\alpha (p_B-k)= \delta(\omega - \omega^\prime) \, 
\bar v_\beta (k) \, u_\alpha (p_B-k),  
\label{eq:phih0} \\
\Phi^{q \bar q^\prime (0)}_{\gamma\delta} (x) 
& = &\phi^{q \bar q^\prime (0)} \,
\bar u_\delta (\Unew p) \, v_\gamma(\bar u p) =\delta(u - x) \,
\bar u_\delta (\Unew p) \, v_\gamma(\bar u p) , 
\label{eq:phim0}
\end{eqnarray}
and~${\cal A}_8^{(0)}$ can be written in the factorized form
\begin{equation}
{\cal A}_8^{(0)} = \Phi^{b \bar q^\prime (0)} 
\star {\cal T}_8^{{\rm II}(0)} \star \Phi^{q \bar q^\prime (0)},
\label{eq:A80-convolution}
\end{equation}
with
\begin{eqnarray}
{\cal T}^{{\rm II}(0)}_{8,\, \alpha\beta\gamma\delta} (\omega, u) &=&
\frac{\overline m_b}{m_b}
\frac{C_F}{N_c}\frac{e Q_d \alpha_s}{\pi} \, 
\frac{1}{\Unew\omega}  \left\{
{\epsilon\llap{/}}_{\perp} \gamma_{\nu_\perp} (1 + \gamma_5)  
\right\}_{\delta\alpha} \, 
\left\{ \gamma^{\nu_\perp} \right\}_{\beta\gamma} \nonumber
 \\
&\equiv & t_8^{{\rm II}(0)}\left\{
{\epsilon\llap{/}}_{\perp} \gamma_{\nu_\perp} (1 + \gamma_5)  
\right\}_{\delta\alpha} \, 
\left\{ \gamma^{\nu_\perp}  \right\}_{\beta\gamma}.
\label{eq:t80} 
\end{eqnarray}
The sub-factorization of $t_8^{{\rm II}(0)}$ 
into the convolution of a hard coefficient with the jet function is 
given by  \cite{Becher:2005fg}
\begin{equation}
\label{eq:leadingC}
\Delta_8 C^{B1(0)}(\tau)=-\frac{ eQ_d {\overline m}_b}{4\pi^2}\frac{\bar\tau}
{\tau}; \qquad 
j_\perp^{(0)}(\tau; u,\omega) =-\frac{4\pi C_F\alpha_s}{N_c}
 \frac{1}{m_b \omega \bar u}\delta(\tau-u).
\end{equation}
To show that the hard-scattering kernel $t_8^{\rm II}$ is what appears
in the factorization formula (\ref{eq:SCETff}), we now consider
in more detail the hadronic matrix elements of four-quark
operators in ${\rm SCET}_{\rm II}$.  Note that the four-quark
operator whose hadronic matrix element leads to a product
of LCDAs has the opposite Fierz ordering ($[\bar \xi \xi ] [\bar q_s h_v]$)
compared to the operator whose partonic matrix element matches
straightforwardly onto the expression in (\ref{eq:A80-convolution}) 
($[\bar \xi h_v ] [\bar q_s \xi]$). In four dimensions the two 
operators are connected by a Fierz transformation according to
(see, e.g.,  \cite{Beneke:2005gs})
\begin{equation}
O_{V_\perp}=\bar \xi 
{\epsilon\llap{/}}_{\perp} \gamma_{\nu_\perp} (1 + \gamma_5) h_v \,
\bar q_s \gamma^{\nu_\perp} \xi \quad \leftrightarrow \quad 
O^\prime_{V_\perp} =
\bar \xi {\epsilon\llap{/}}_{\perp}\frac{\nslash_+}{2}
(1+\gamma_5)\xi \,
\bar q_s \, \frac{\nslash_-}{2} \gamma_5 \,h_v \, ,
\label{eq:Fierz}
\end{equation}
where the collinear (soft/HQET) fields in each operator are understood to 
be evaluated at different points on the $n_+ (n_-)$  light-cone, 
and made gauge invariant by inserting  appropriate  Wilson lines. 
In writing (\ref{eq:Fierz}) we have omitted Dirac structures contributing
to  $O^\prime_{V_\perp}$ which have no projection onto the pseudoscalar
$B$-meson LCDA.  Comparing with (\ref{eq:HadLCDAs}), we immediately see that,
in the absence of soft-collinear interactions, the hadronic
matrix element of the operator $O_{V_\perp}^\prime$ factorizes into a product
of $\phi_\perp^V$ and $\phi^B_+$.  On the other hand, 
the partonic matrix element of the Fierz-transformed
version $O_{V_\perp}$ matches (\ref{eq:A80-convolution}), up
to the hard-scattering kernel $t_8^{\rm II(0)}$.  We thus
verify that our expression for the tree-level amplitude
is equivalent to (\ref{eq:SCETff}).

We shall perform a similar calculation at one loop in the 
next subsection. A  complication compared to  
tree level is that the appearance of IR poles in the 
dimensionally regulated one-loop amplitude prevents one from using 
the Fierz relation (\ref{eq:Fierz}).  To extract the hard-scattering
kernel by comparing the renormalized matrix elements calculated in the 
two Fierz orderings is thus non-trivial, and  will require the
use of some technical details from the SCET analysis used to extract
the one-loop jet function in  \cite{Beneke:2005gs,Becher:2004kk}.

\subsection{$Q_8$ at one loop} 
\label{sec:NLO-contribution} 

\begin{figure}
\begin{center}
\includegraphics[width=0.9\textwidth]{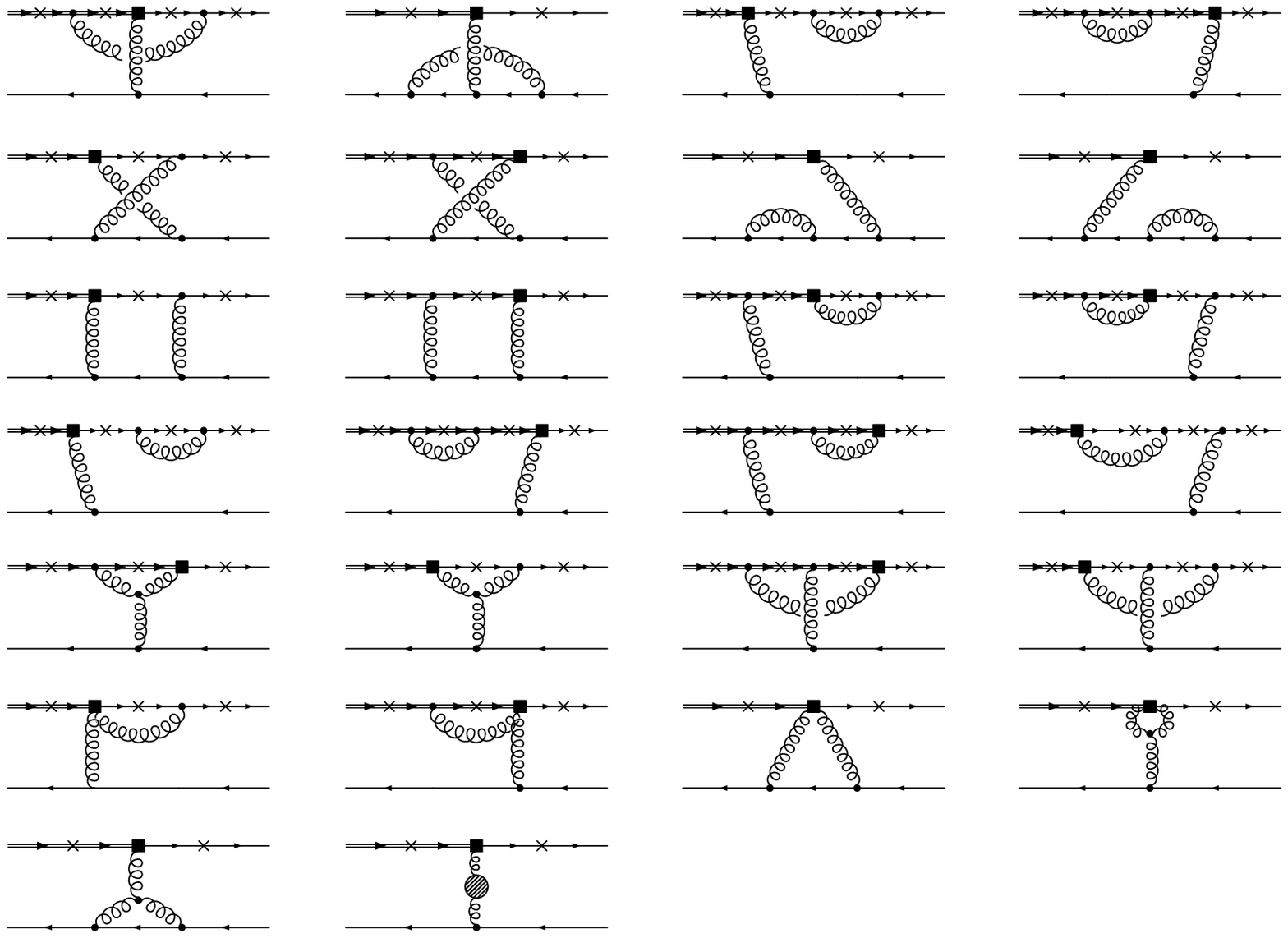}
\end{center}
\vspace{-.5cm}
\caption{The one-loop corrections to spectator scattering with
$Q_8$.  The solid box denotes a $Q_8$ insertion and the photon can 
be attached to any of the crosses.  }
\label{fig:oneloop}
\end{figure}

We now turn to a  main subject of this paper, 
the calculation of the one-loop correction from $Q_8$
to the hard-scattering kernel $t^{\rm II}$.  
The first task is to calculate the amputated part of the full 
set of one-loop Feynman diagrams shown in Figure \ref{fig:oneloop},
supplemented by the on-shell renormalization factors for the quark
fields.  Photon emission from the spectator quark need not 
be considered for the 
case of pseudoscalar $B$-meson decay, since the four-quark 
structures appearing in the matching vanish at leading order in
$1/m_b$ after projecting onto the meson LCDAs \cite{Becher:2005fg}.  
When a loop integral involves more than one scale, we calculate 
the leading term in the $1/m_b$ expansion using the method of regions.
All integrals are calculated in dimensional regularization in
$d=4-2\epsilon$ dimensions and with the NDR scheme for $\gamma_5$.  
The result can be written in the form
\begin{equation}
{\cal A}_8^{(1)}= A^{(1)} \left({\epsilon\llap{/}}_\perp \gamma_{\nu_\perp} 
\otimes  \gamma^\nu_{\perp}\right) + B^{(1)} \left(\gamma_{\nu_\perp} 
{\epsilon\llap{/}}_\perp 
\otimes  \gamma^\nu_{\perp} \right).
\label{eq:OneLoopA}
\end{equation}
The four-quark structure multiplied by the scalar function $A^{(1)}$
is related to the matching of the operator 
$J^{B1}$. It is proportional to the tree amplitude and is 
used to extract the hard-scattering kernel.
For our choice of external momenta, the function
$A^{(1)}$  receives non-vanishing contributions 
from the hard, hard-collinear, and soft momentum regions.  
Since we work with on-shell partonic states, 
contributions from the collinear and soft-collinear regions 
are only in scaleless integrals and vanish 
(this would not be true if off-shell 
regularization were used).
We shall label the contributions from the different regions as 
$A^{(1)}_{h}, A^{(1)}_{hc}$, and $A^{(1)}_{s}$
in what follows. Moreover,  we define the amplitude $A^{(1)}_h$ to include
the $\alpha_s$ contribution from wave-function renormalization of 
the $b$-quark field, which reads
\begin{equation}
Z_{2b}^{1/2}-1=-\frac{C_F\alpha_s}{4\pi}\left(\frac{3}{2\epsilon}
+3\ln\frac{\mu}{m_b}+2\right) \,.
\end{equation}
The renormalization factors for the light-quark fields vanish,
to this order in $\alpha_s$.
In writing the result (\ref{eq:OneLoopA}), we used
the prescription \cite{Beneke:2005gs}
\begin{eqnarray}
\label{eq:ev}
\gamma^{\rho_\perp}\gamma^{\lambda_\perp} 
{\epsilon\llap{/}}_\perp \gamma^{\mu_\perp}\otimes \gamma_{\mu_\perp} 
\gamma_{\lambda_\perp}\gamma_{\rho_\perp} &\to& (d-4)^2 
\left({\epsilon\llap{/}}_\perp \gamma^{\nu_\perp} 
\otimes  \gamma_{\nu_\perp}\right) ,
\end{eqnarray}
the relevance of which will be explained below.

The structure multiplied by the scalar function $B^{(1)}$ 
is related to the matching of the operator
$J^{B2}$. The contributions from individual diagrams contain
$1/\epsilon$ poles, but these cancel in the sum of all diagrams. 
Since the matrix element of this four-quark operator 
has no projection onto the $B$-meson LCDA, this piece does not
contribute to the hard-scattering kernel.

Note that there is no third Dirac structure, which would correspond
to a contribution from the operator $J^{A}$.  The matching of $J^{A}$
involves the emission of the $n_+A_{hc}$ component of a hard-collinear
gluon. Using the equations of motion 
$\bar v(k)\nslash_+= \nslash_-  v(\bar u p)=0$ ensures that the 
Dirac structure can always be written in terms of the 
transverse components of Dirac matrices.  It is easy to see that
this precludes taking out a factor of $n_+A_{hc}$ from the 
Wilson line in the operator $J^A$ and attaching it to the spectator quark.

The function $A^{(1)}$ has both UV and IR divergences.    
The UV divergences are removed by coupling constant, 
mass, and operator renormalization (recall that the contribution from 
wave-function renormalization of the $b$-quark is included in the definition
of $A^{(1)}_h$). 
We define the renormalized parameters as $\overline{m}_b^{\rm bare}=Z_m 
\overline{m}_b$, etc.  As with the vertex term,  
we first compute the QCD amplitude
with $n_f=n_h+n_l$ active flavors, and then express results 
in the $\MSbar$ scheme in the $n_l$-flavor theory by renormalizing 
the coupling as in (\ref {eq:Znhnl}).  Doing so, we find that
all dependence on $n_h$ drops out.  The UV renormalized amplitude 
is obtained by making the replacement
\begin{eqnarray}
 A^{(1)}&\to &  A^{(1)}+A^{(1)}_{\rm c.t.}= A^{(1)}+
\left(Z_\alpha^{n_h+n_l(1)}+Z_m^{(1)}+Z_{88}^{(1)}- \frac{\Unew}{\bar u}
\frac{Z_{87}^{(1)}}{Q_d} \right)A^{(0)},
\end{eqnarray}
where the various factors at one loop are
\begin{eqnarray}
Z_m^{(1)}=-\frac{3C_F \alpha_s}{4\pi\epsilon}\,,\quad
Z_{88}^{(1)}=8\left(C_F-\frac{C_A}{4}\right)\frac{\alpha_s}{4\pi \epsilon}\,,
\quad 
Z_{87}^{(1)}=\frac{ Q_dC_F \alpha_s}{\pi\epsilon}
\label{eq:renfacts}\,.
\end{eqnarray} 
The $u$-dependent factor multiplying the counterterm
$Z_{87}^{(1)}$ follows from the tree-level coefficients 
(\ref{eq:leadingC7},\ref{eq:leadingC}). To this order 
in $\alpha_s$, mass renormalization for the $b$-quark 
is needed only for the $\MSbar$ mass appearing in the definition
of $Q_8$, see (\ref{eq:penguin-operators}).

We can extract the one-loop correction to 
the hard-scattering kernel  from the UV renormalized 
partonic amplitude.
It is defined by 
\begin{equation}
\phi^{b \bar q^\prime (0)} \star 
 t_8^{{{\rm II}} (1)} \star
\phi^{q \bar q^\prime (0)} =  A^{(1)}+A^{(1)}_{\rm c.t.} - 
\phi^{b \bar q^\prime (1)} \star 
t_8^{{{\rm  II}} (0)} \star
\phi^{q \bar q^\prime (0)} - 
\phi^{b \bar q^\prime (0)} \star
t_8^{{{\rm  II}} (0)} \star
\phi^{q \bar q^\prime (1)} ,
\label{eq:t2def} 
\end{equation} 
where the one-loop LCDAs are the renormalized ones.
The one-loop contributions to the renormalized LCDAs take the form
\begin{eqnarray}
\label{eq:renLCDAs}
\phi^{q \bar q^\prime(1)}&=&Z_{V_\perp}^{(0)} 
\star \phi^{q \bar q^\prime(1)}_{\rm bare}+Z_{V_\perp}^{(1)} 
\star \phi^{q \bar q^\prime(0)}_{\rm bare}, \\
\phi^{b \bar q^\prime(1)}&=&Z_B^{(0)}
\star \phi^{b \bar q^\prime(1)}_{\rm bare}+
Z_B^{(1)} \star \phi^{b \bar q^\prime(0)}_{\rm bare}.
\end{eqnarray}
The renormalization factor $Z_{V_\perp}$ for 
the $V$-meson LCDA is the Brodsky-Lepage kernel 
\cite{Lepage:1980fj, Efremov:1979qk} for a transversely polarized
vector meson, and that for the  $B$-meson was calculated 
in \cite{Lange:2003ff}.  Here there is an important subtlety, which 
is discussed in detail in \cite{Beneke:2005gs,Becher:2004kk}.
The renormalization factors for the $B$ and $V$ meson LCDAs
are calculated in the $\MSbar$ scheme
under the assumption that the light-cone projection 
operators in (\ref{eq:phihdef}, \ref{eq:phimdef}) are contracted with the
specific Dirac structures shown in (\ref{eq:HadLCDAs}).  In our
case this corresponds to the one-loop corrections to the operator
$O^\prime_{V_\perp}$ in (\ref{eq:Fierz}).  However, the one-loop
amplitude $A^{(1)}$ is extracted from (\ref{eq:OneLoopA}) and 
thus multiplies the opposite Fierz ordering, corresponding
to $O_{V_\perp}$ in (\ref{eq:Fierz}).  The Fierz transformation
relating the two operators is valid only in $d=4$ dimensions, and 
in $d=4-2\epsilon$ dimensions they may differ by  terms which
vanish in the limit $d \to 4$.  One can fix these seemingly arbitrary 
terms by defining a prescription for the reduction of evanescent Dirac
structures, which ensures that the subtractions
in (\ref{eq:t2def}) are performed according to the $\MSbar$ scheme.
This is achieved by using the prescription for the evanescent operator in 
(\ref{eq:ev})  \cite{Beneke:2005gs,Becher:2004kk}.
Since this  structure is multiplied by poles related to the 
hard-collinear region,  we can verify that we have performed the 
correct subtractions by checking eq. (\ref{eq:Jconv}) 
against a corresponding result obtained using the one-loop
jet function from \cite{Beneke:2005gs,Becher:2004kk}.  
We describe this cross-check below.

To evaluate (\ref{eq:t2def}) it is instructive to rewrite 
the right-hand side as
\begin{eqnarray}
\label{eq:renT2}
(\ref{eq:t2def})&=& A^{(1)}_{h+hc}+A^{(1)}_{\rm c.t.} -
Z_B^{(1)}  \star 
t_8^{{\rm II} (0)} \star
\phi^{q \bar q^\prime (0)} - 
\phi^{b \bar q^\prime (0)} \star
t_8^{{\rm II} (0)} \star
Z_{V_\perp}^{(1)} \nonumber \\
&&+ A^{(1)}_{s}- \phi^{b \bar q^\prime(1)}_{\rm bare} \star 
t_8^{{\rm II} (0)} \star\phi^{q \bar q^\prime (0)}.
\end{eqnarray}
We have simplified the above equation by using
that the factors $Z_i^{(0)}$ and $\phi^{i(0)}$ are delta functions,
and that the one-loop correction to the bare
$V$-meson LCDA vanishes for on-shell quarks in dimensional
regularization (because the integrals are scaleless).
Using the explicit results from 
\cite{Lepage:1980fj, Efremov:1979qk, Lange:2003ff} to
perform the convolutions with the tree-level hard-scattering
kernel, the first line of (\ref{eq:renT2}) can be written as  
\begin{equation}
\label{eq:IRsubtractions}
 A^{(1)}_{h+hc}+A^{(1)}_{\rm c.t.}
-\left(z_{B}^{(1)}+z_{V_{\perp}}^{(1)}\right)A^{(0)},
\end{equation}
where 
\begin{eqnarray}
z_{B}^{(1)}&=&\frac{C_F \alpha_s}{4\pi}
\left[\frac{1}{\epsilon^2}+
\frac{2}{\epsilon}\ln\frac{\mu}{\omega}-\frac{5}{2\epsilon}\right],
\nonumber \\
z_{V_\perp}^{(1)}&=& \frac{C_F \alpha_s}{4\pi\epsilon}
\left[-3-\frac{2 \ln \Unew}{\bar u}\right].
\end{eqnarray}
Evaluating (\ref{eq:IRsubtractions}), 
one finds that after UV renormalization 
the IR poles in the sum of the hard and 
hard-collinear regions are exactly subtracted
by the poles related to the renormalization of the LCDAs.  
On the other hand, the second line of (\ref{eq:renT2})
vanishes, showing that contributions from the soft region 
are restricted to the $B$-meson LCDA. Therefore, 
the hard-scattering kernel is free of 
$1/\epsilon$ poles and insensitive to IR physics.
This was first  verified by the explicit calculations in 
\cite{Descotes-Genon:2004hd}.

We have  shown that the finite part of the sum of 
hard and hard-collinear regions is $t_8^{{\rm II}(1)}$.  
As for $Q_7$, we write the result as 
\begin{equation}
t^{{\rm II}(1)}_8=  \Delta_8C^{B1(0)}\star j_\perp^{(1)}+ 
\Delta_8C^{B1(1)}\star j_\perp^{(0)}.
\end{equation}
We can identify the finite part 
of the hard-collinear region with $\Delta_8 C^{B1(0)}\star j_\perp^{(1)}$, 
and the finite part of the hard region with 
$\Delta_8C^{B1(1)}\star j_\perp^{(0)}$. 
For the sum of hard-collinear graphs, we find
\begin{equation}
\label{eq:Jconv}
\Delta_8C^{B1(0)} \star j_\perp^{(1)}= A^{(1)}_{\rm { hc,fin}}=
\frac{\alpha_s}{(4\pi)}[C_F j_F + C_A j_A + n_l j_f]t_8^{{\rm II}(0)},
\end{equation}
where $n_l=4$ is the number of light flavors, and 
($L_{\rm hc}=\ln m_b\omega/\mu^2)$
\begin{eqnarray}
\label{eq:J}
j_F&=& L_{\rm hc}^2+ \left( 5-\frac{2-2\ln \Unew}{\bar u}+
2\ln \Unew\right)L_{\rm hc}
+\frac{4}{\bar u}-12-\frac{\pi^2}{6} 
\nonumber \\
&&+\left(5-\frac{4}{\bar u}\right)\ln \bar u 
+\frac{2 \ln \bar u \ln \Unew}{\bar u}+\ln^2\Unew \left(1+\frac{1}{\bar u}\right)
 + \frac{4}{\bar u}{\rm Li}_2(\bar u), \nonumber \\
j_A &=& \left(-\frac{11}{3}+\ln \bar u -
\left(1+\frac{1}{\bar u^2}\right) \ln \Unew\right)L_{\rm hc}+\frac{76}{9}
+\left(-\frac{11}{3}+\frac{1}{\bar u}\right)\ln \bar u+\frac{\ln^2 \bar u}{2}, 
\nonumber \\
&& +\left(\frac{1}{\bar u^2}-\frac{1}{\bar u}\right)\ln \Unew
-\left(\frac12+\frac{1}{2\bar u^2}\right)\ln^2 \Unew-\frac{\ln \bar u \ln \Unew}{\bar u^2}
-\frac{2}{\bar u^2}{\rm Li_2}(\bar u), \nonumber \\
j_l&=& \frac{2}{3} L_{\rm hc} + \frac{2\ln \bar u}{3}-\frac{10}{9}.
\end{eqnarray} 
Taking the convolution of the one-loop jet function $j_\perp^{(1)}$ 
listed in the Appendix with the leading-order hard coefficient
in (\ref{eq:leadingC}), 
we reproduce the above equation. This verifies that 
the sub-factorization of the hard-scattering kernel according
to momentum regions is equivalent to that in SCET, and also
that we have performed the correct subtractions  in 
(\ref{eq:t2def}).  However, we again
emphasize that this integrated form cannot be used to 
obtain the resummed hard-scattering kernels used in our numerical
analysis in Section \ref{sec:numerics}.

The finite part of the hard region gives an expression for 
$\Delta_8 C^{B1(1)}\star j_\perp^{(0)}$. In this case
we have
\begin{equation}
\Delta_8C^{B1(1)}\star j_\perp^{(0)}=A^{(1)}_{{\rm h,  fin}} =
\frac{\alpha_s}{4\pi}[C_F h_F + C_A h_A]t_8^{{\rm II}(0)}\, ,
\end{equation}
where
\begin{eqnarray}
\label{eq:hardC8}
&&h_F=-2L^2 -\left(1+\frac{4}{\bar u}-4 \ln\Unew\right)L
-\left(18-\frac{8}{\bar u}\right)L_{\rm QCD}
+ i \pi \left(-4+\frac{2}{\bar u}+\frac{2\ln\Unew}{\bar u^2}\right) \nonumber \\
&&
-15+\frac{7\pi^2}{12}-\frac{1}{\bar u}-\frac{2\pi^2}{3\bar u} 
 +\left(2+\frac{2}{\Unew}+\frac{4}{\bar u}\right)\ln \bar u
\nonumber \\ &&
+ \left(4-\frac{2}{\bar u}-\frac{2}{\bar u^2}\right)\ln\Unew
+\left(\frac{3}{\bar u}+\frac{2}{\bar u^2}+\frac{1}{2-\bar u} \right)\ln \bar u \ln\Unew
-\left(2+\frac{1}{\bar u^2}\right)\ln^2 \Unew
\nonumber \\
&&+ \left(-2+\frac{1}{2\bar u}+\frac{3}{\bar u^2}+\frac{1}{2(2-\bar u)}\right){\rm Li}_2(\bar u)
+ \left(\frac{5}{\bar u}-\frac{6}{\bar u^2}+\frac{1}{2-\bar u}\right)g(\bar u)
+\left(\frac{1}{\bar u}-\frac{1}{2-\bar u}\right)h(\bar u), \nonumber \\
&& \nonumber \\
&& 
h_A= \left(-2\ln\Unew - \frac{2 \ln\Unew}{\bar u^2}+2\ln \bar u\right)L+4L_{\rm QCD}
+ i\pi\left(1-\frac{1}{\bar u}-\frac{2\ln\Unew}{\bar u^2}\right)
+2-\frac{\pi^2}{3}+\frac{3}{\bar u} \nonumber \\
&& -\frac{\ln \bar u}{\bar u}+\left(-1+\frac{2}{\bar u}-\frac{1}{\bar u^2}\right)\ln \Unew
+\left(1-\frac{3}{2\bar u}-\frac{1}{2(2-\bar u)}\right)\ln \bar u \ln \Unew
-\ln^2 \bar u \nonumber\\
&&+\left( 1+\frac{1}{\bar u^2}\right)\ln^2 \Unew  + \left(2-\frac{1}{4\bar u}-\frac{1}{2\bar u^2}-\frac{1}{4(2-\bar u)}\right){\rm Li}_2(\bar u)
\nonumber \\ 
&&
+\left(-\frac{5}{2\bar u}+\frac{3}{\bar u^2}-\frac{1}{2(2-\bar u)}\right)g(\bar u)
+\left(-\frac{1}{2\bar u}+\frac{1}{2(2-\bar u)}\right) h(\bar u). 
\end{eqnarray}
The logarithms $L$ and $L_{\rm QCD}$ are defined after (\ref{deltac7a}).
The terms proportional to $L$ agree with the corresponding
terms in \cite{Descotes-Genon:2004hd}, and can be deduced by
convoluting the renormalization factor
$z_\perp(\tau)$ of the SCET current $J^{B1}$ obtained
in \cite{Beneke:2005gs, Hill:2004if} with the tree-level
hard coefficient $\Delta_8 C^{B1}$ in (\ref{eq:leadingC}).  
We have defined the functions
\begin{eqnarray}
g(u)&=&\int_0^1 dy\, \frac{\ln\left[1-u y(1-y)\right]}{y}, \\
h(u)&=&\int_0^1 dy \, \frac{\ln\left[1-u y(1-y)\right]}{1-u y}.
\end{eqnarray}
These functions have no imaginary part for $u\in [0,1]$ and can  
be expressed in terms of dilogarithms and logarithms, but we 
shall not give the explicit results here.  Since $j_\perp^{(0)}$ 
is a delta function in $\tau$, the result for 
$\Delta_8 C^{B1(1)}$ is obtained directly from (\ref{eq:hardC8}).

\section{Numerical analysis}
\label{sec:numerics}
In this section we discuss the numerical impact of our results.  
Our main focus is on the branching fractions for 
$B\to K^*\gamma$ and $B_s\to \phi \gamma$ decays. 
The branching fraction for $B\to K^*\gamma$ decays is 
\begin{equation}
\label{eq:BF}
{\cal B}(B\to K^*\gamma)=\frac{\tau_B m_B}{4\pi}\left(1-
\frac{m_{K^*}^2}{m_B^2}\right)\left | {\cal A}_{\rm v}+ {\cal A}_{\rm hs}
\right|^2\, ,
\end{equation}
where we have split the contributions from the vertex and 
hard-spectator corrections according to 
\begin{eqnarray}
{\cal A}_{\rm v}&=&\frac{G_F}{\sqrt 2} V_{cs}^* V_{cb}
\sum_i C_i(\mu_{\rm QCD}) \, \Delta_i C^{A}(m_b,\mu_{\rm QCD},\mu) 
\zeta_{K_\perp^*}(\mu), \\
{\cal A}_{\rm hs}&=&
\frac{G_F}{\sqrt 2} V_{cs}^* V_{cb}
\sum_i C_i(\mu_{\rm QCD})  \,t_i^{{\rm II}}(\mu_{\rm QCD},\mu)
\star \left(\frac{\sqrt{m_B} F}{4} 
\phi^B_+ \star f_{K^*} \phi_{K^*_{\perp}}\right)(\mu) .
\end{eqnarray}
Results for $B_s\to \phi \gamma$ are obtained by making
the appropriate replacements.  The branching fractions
depend on a number of parameters, whose values and uncertainties are 
summarized in Table~\ref{tab:input-param}.
The vertex and hard-spectator amplitudes are independently 
invariant under variations of $\mu_{\rm QCD}$ and $\mu$.
For this reason, their contributions to the amplitudes can
be studied separately.  We  discuss each one in turn for 
the case of $B\to K^*\gamma$, before presenting the final branching 
fractions for all decay modes in Section \ref{sec:BranchingFractions}. 

\subsection{The vertex amplitude}
\label{sec:VertexAmplitude}
We begin with the vertex corrections.  For these corrections
the relevant perturbative quantities are the
Wilson coefficients $C_i$ in the effective weak Hamiltonian 
and the SCET matching coefficients $\Delta_i C^{A}$.
We calculate the Wilson coefficients in the effective weak Hamiltonian 
using  the information summarized in Appendix A of 
\cite{Czakon:2006ss} (see also \cite{Bobeth:1999mk,Misiak:2004ew }), 
and collect the results 
for three values of the renormalization scale in  
Table~\ref{tab:wilson-coeff}.  
The results for the SCET matching coefficients
are simplest when $\mu_{\rm QCD}= \mu = m_b$, in which case
all logarithms vanish. We use this choice as our  default scheme. 
Throughout the analysis we use the four-loop running 
coupling with $\alpha_s(m_Z)=0.1176$, switching from
five to four active flavors at the matching scale 
$\mu_h=\mu_{\rm QCD}$.  The vertex amplitude and the 
branching fractions depend rather strongly on $\zeta_{V_\perp}$.
We determine it in Section \ref{sec:SoftFunction}
by requiring that the matrix element of $Q_7$ be proportional to the 
QCD form factor $F^{B\to V_\perp}$ at NNLO, finding  
$\zeta_{V_\perp}(\mu=m_b)= 0.35\pm 0.05$.  
We will express higher-order corrections to the amplitudes
in terms of the leading-order result, which is
\begin{equation}
{\cal A}_{\rm v}^{\rm LO} = -\frac{G_F}{\sqrt{2}}V_{cs}^* V_{cb}
C_7^{LL}\,\frac{e\,{\overline m}_b \,2E_\gamma}{4\pi^2}
\zeta_{K^\star_\perp}
=-5.48 \times 10^{-9} \,.
\end{equation}
Up to NNLO, the result obtained using the default parameter values 
in Table~\ref{tab:input-param} is
\begin{equation}
\frac{{\cal A}_{\rm v}^{\rm NNLO}}{{\cal A}_{\rm v}^{\rm LO}}=
1+\left(0.096+0.057 i\right)\left[\alpha_s\right]
+ \left(-0.007+0.030 i\right)\left[\alpha_s^2\right],
\end{equation}
where the first term in parentheses is the NLO ($\alpha_s)$
correction and the second term the NNLO  $(\alpha_s^2)$ 
correction.  The corrections come both from the 
Wilson coefficients $C_i$ in the effective weak
Hamiltonian and the SCET coefficients $\Delta_i C^{A}$.
Note that we have used the effective coefficients $C_{7,8}^{\rm eff}$
in the numerical analysis.  This amounts to including 
certain contributions from $Q_3\dots Q_6$, and should be 
taken into account, if in the future the contributions from 
these operators are worked out systematically.
Split into contributions from the individual operators, the 
results for the NLO and NNLO perturbative corrections read
\begin{eqnarray}
\frac{{\cal A}_{\rm v}^{\rm NNLO}}{{\cal A}_{\rm v}^{\rm LO}}-1
&=& 
\bigg((0.264+  0.034i)\, [Q_1] -(0.184) \,[Q_7] 
+(0.016+0.023i) \,[ Q_8] \bigg)\,[\alpha_s]
 \\
&+& \bigg((0.073+0.022i)\, [Q_1] -(0.081) \,[Q_7] 
+(0.002+0.008i) \,[ Q_8] \bigg)\,[\alpha_s^2] \nonumber \,.
\end{eqnarray}
At both NLO and NNLO the corrections from
$Q_1$ and $Q_7$ are relatively large, but the real parts
tend to cancel against each other. Whether this cancellation persists beyond
the large-$\beta_0$ limit is an important question.
The contribution from $Q_8$ to the real part of the amplitude 
is small, but that to the imaginary part is not.  It adds together
with that from $Q_1$ to produce a large NNLO correction to the
imaginary part.  This would be a significant effect for CP asymmetries,
a topic we leave for future work.

In Section \ref{sec:BranchingFractions} we will study the 
dependence of the branching fractions on the renormalization
scales.  To do this we use the SCET Wilson coefficients as 
given in (\ref{eq:CAev}) in Section \ref{sec:Vertex}.  As explained there, 
this allows us to fix the scale $\mu=m_b$ in 
$\zeta_{V_\perp}$ and study the stability
of the results under variations in $\mu_{\rm QCD}$ and $\mu_h$.
Although the expressions in the Appendix allow us to vary $\mu_{\rm QCD}$
and $\mu_h$ separately, we choose not to do so.  For simplicity,
we set $\mu_{\rm QCD}=\mu_h$ and vary them simultaneously. 
To evaluate the RG exponents in the SCET evolution factors
we distinguish the operators $Q_{7,8}$ and $Q_1$.  
For $Q_{7,8}$ we evaluate the RG exponents
using the two-loop anomalous dimensions in $a$ and $a_J$, and
the three-loop cusp anomalous dimension in the Sudakov
factor $S$. For $Q_1$ we evaluate the RG exponents using
the large-$\beta_0$ limit.  In that case it is consistent
to set all SCET anomalous dimensions to zero, meaning that 
we can use the form (\ref{eq:c1Annlo}) directly.

\begin{table}[t]
\caption{
Wilson coefficients $C_i(\mu)$ $(i=1,7,8)$ at LL, NLL and NNLL.
The results at NNLL are calculated from the expressions given in
\cite{Bobeth:1999mk,Misiak:2004ew, Czakon:2006ss}, adapted to the 
operator basis  in (\ref{eq:4-quark-operators}, \ref{eq:penguin-operators}).
The table uses $m_b=4.8$~GeV.}
\label{tab:wilson-coeff}
\begin{center} 
\begin{tabular}{|c|c|c|c|}
\hline 
& LL & NLL & NNLL \\ \hline
$C_1 (\mu=m_b)$ & $  1.11$ & 1.06&   \\ 
$C_1(\mu=\sqrt{2}m_b)$   & $1.09$ & $1.04$&  \\
$C_1(\mu= m_b/\sqrt{2})$ & $1.13$ & $1.08$&  \\ \hline 
$C_7^{\rm eff}(\mu=m_b)$ & $ -0.312$ &$ -0.303$ &$ -0.294$ \\
$C_7^{\rm eff}(\mu=\sqrt{2}m_b)$  & $-0.294$  & $-0.290$ & $ -0.282$ \\
$C_7^{\rm eff}(\mu= m_b/\sqrt{2})$& $-0.332 $  & $-0.316$ & $ -0.306$ \\ \hline
$C_8^{\rm eff} (\mu=m_b)$ & $-0.148$ & $-0.167$ & \\
$C_8^{\rm eff}(\mu=\sqrt{2}m_b)$   & $-0.141$ & $-0.159$&  \\
$C_8^{\rm eff}(\mu= m_b/\sqrt{2})$ & $-0.156$ & $-0.175$&  \\ 
 \hline 
\end{tabular}
\end{center}
\end{table}

\subsection{The hard spectator amplitude}
\label{sec:SpectatorAmplitude}
The evaluation of the hard spectator amplitude is 
more complicated than the vertex amplitude. It 
involves a large number of hadronic parameters and 
the hard-scattering kernel contains logarithms of both 
the hard and hard-collinear scales.  While it is possible
to fix the scale $\mu_{\rm QCD}\sim m_b$ to eliminate some
of these logarithms, any choice of the SCET factorization scale 
$\mu$ leads to large logarithms in $t^{{\rm II}}_i$.  
This can be solved by renormalization-group improvement 
in the effective theory \cite{Hill:2004if}. The hard coefficient
$\Delta_i C^{B1}$ is extracted at a scale 
$\mu_h\sim \mu_{\rm QCD}\sim m_b$ and 
and evolved down to the intermediate scale $\mu_i\sim 1.5$~GeV by
solving the RG equations in the effective theory.  
The RG-improved hard coefficients read \cite{Hill:2004if}
\begin{equation}
\label{eq:resummedCB}
\Delta_i C^{B1}(u,\mu_i) =
\left(\frac{m_b}{\mu_h}\right)^{a(\mu_h,\mu_i)} e^{S(\mu_h,\mu_i)}
\int_0^1 dv \,U_\perp(u,v,\mu_h,\mu_i) \Delta_i C^{B1}(v,\mu_h)\, .
\end{equation}
The RG exponents $S$ and $a$ are the same as in 
(\ref{eq:CuspExps}).
The  evolution factor $U_\perp$ is the solution to the
integro-differential equation 
\begin{equation}
\mu \frac{d}{d\mu} U_\perp(u,v,\mu_h,\mu)
=\int_0^1 dy \, \gamma_\perp(y,u) U_\perp(y,v,\mu_h,\mu),
\end{equation}
with the initial condition $U_\perp(u,v,\mu_h,\mu_h)=\delta(u-v)$.
The distribution $\gamma_\perp(y,u)$ is the
anomalous dimension of the operator  $J^{B1}$.
A proper treatment of the NNLO matching corrections
requires this anomalous dimension at two loops, but  
at present it is known only at one loop 
\cite{Beneke:2005gs, Hill:2004if}.  This adds a small
uncertainty to the analysis. The solution to the evolution 
equation is obtained  numerically.  In the numerical 
implementation we perform the $\mu$-evolution from 
$\mu_h$ to $\mu_i$ in 100 discrete steps.   We choose the 
default renormalization scales as $\mu_{\rm QCD}=\mu_h=m_b$ 
and $\mu_i=1.5$ GeV.  
The dependence on the variable $u$ in the resummed $\Delta_i C^{B1}$
is obtained for discretized values of $0<u<1$.  We 
determine the discretization scale by taking more points
in $u$ until the numerical convolution of the resummed coefficient 
with the jet function becomes stable.
This generally requires between one and three-hundred values, 
although for some cases it is necessary to take more values 
near the endpoints. 

It is natural to evaluate the resummed hard coefficients 
$\Delta_i C^{B1}(u,\mu)$ at a scale $\mu \sim \mu_i$, since
at that scale the jet function is free of large logs.  However, 
the hadronic parameters in  Table~\ref{tab:input-param} 
are extracted at a low scale $\mu=1$ GeV.  For a proper treatment one 
must either run these parameters up to the intermediate scale
$\mu_i\sim 1.5$ GeV, or run the hard-scattering kernel down
to the lower scale.  This stage of RG running has been studied 
in \cite{Hill:2004if,Bosch:2003fc,Lange:2003ff}.  
We have performed this evolution 
in our numerical analysis but its effect on the branching fractions
is extremely small. Therefore, in quoting our results, 
we perform the running  from $\mu_h$ to $\mu_i$, 
but ignore that between the scale $\mu_i$ and the factorization scale
$\mu\sim 1$ GeV. The shortcoming of this treatment is that the amplitude
is not invariant under variations of the intermediate scale.
However, the dominant effect in this scale variation is 
related the $B$-meson distribution amplitude.   
We account for this in our error analysis by assigning a 
rather large uncertainty to $\lambda_B$.

A complete treatment of the hard-spectator amplitude
is only possible for the NLO  corrections. 
There are three pieces missing for a full resummed 
result for the hard-spectator term at NNLO:  
the NNLO hard matching coefficient for $Q_1$, 
the two-loop anomalous dimension of the current $J^{B1}$, and
the two-loop anomalous dimension of the jet function. 
These missing pieces add uncertainties to the analysis 
which are difficult to quantify.
However, we will see that the higher-order corrections from 
spectator scattering are not very important for the branching
fractions.  

In addition to the input parameters listed in 
Table~\ref{tab:input-param},   we must also specify the 
meson LCDAs.  For the vector mesons we use the Gegenbauer
expansion and keep only the first two moments:
\begin{equation}
\phi_V(u)= 6 u(1-u) 
\left[1+ a_1^V(\mu) C_1^{(3/2)}(2u-1)+a_2^V(\mu) C_2^{(3/2)}(2u-1)\right].
\end{equation}
For the $B$-meson LCDA we use the model \cite{Braun:2003wx}
\begin{equation}
\phi_+^B(\omega,\mu= 1\,{\rm GeV})=
\frac{4\lambda_B^{-1}}{\pi}\frac{\omega \, \mu}{\omega^2+\mu^2}
\left[\frac{\mu^2}{\omega^2+\mu^2}
-\frac{2(\sigma_B-1)}{\pi^2}\ln\frac{\omega}{\mu}\right].
\end{equation}
The $B$-meson decay constant in the static limit is
\begin{equation}
F(\mu)= \frac{f_B \sqrt{m_B}}{K(\mu_h)} e^{V_F(\mu_h,\mu)}\,,
\end{equation}
where to one loop \cite{Ji:1991pr}
\begin{equation}
K_F(\mu)=1+\frac{C_F\alpha_s(\mu)}{4\pi}
\left(3\ln \frac{m_b}{\mu}-2\right), \hspace{.7 cm}
V_F(\mu_h,\mu)=-\frac{3C_F}{2\beta_0}
\ln\frac{\alpha_s(\mu)}{\alpha_s(\mu_h)}\,.
\end{equation}

We now quote the result for the hard-spectator
amplitude to NNLO, accurate within the limitations  
explained above.  We find
\begin{equation}
\frac{ {\cal A}_{\rm hs}^{\rm NNLO}}{{\cal A}_{\rm v}^{\rm LO}}=
\big(0.11+0.05 i\big)\,[\alpha_s]+
\big(0.03+0.01 i\big)\,[\alpha_s^2] .
\end{equation}
Performing the RG evolution of the hard-scattering
kernel between $\mu_i$ and the factorization scale $\mu=1$ GeV
suppresses the above result by about $10\%$, or in other words 
makes about a $1\%$ difference on the total amplitude.
Split into contributions from the individual operators, we have 
\begin{eqnarray}
\frac{{\cal A}_{\rm hs}^{\rm NNLO}}{{\cal A}_{\rm v}^{\rm LO}}
&=& 
\bigg((0.023+  0.046i)\, [Q_1] + 0.074 \,[Q_7] 
+0.010 \,[ Q_8] \bigg)\,[\alpha_s]
 \\
&+& \bigg((0.004+0.003i)\, [Q_1] +0.025 \,[Q_7] 
+(0.003+0.005i) \,[ Q_8] \bigg)\,[\alpha_s^2] \nonumber \,.
\end{eqnarray}
Unlike the case of the vertex corrections,
the individual contributions from the different operators 
are rather small at NLO and especially NNLO.
For $Q_1$ we have listed the NNLO correction found 
by numerically evaluating $\Delta_1 C^{B1(0)}\star j_\perp^{(1)}$.  
In addition to this correction from the jet function, there is also
a hard correction $\Delta_1 C^{B1(1)}\star j_\perp^{(0)}$
which is not known. Both terms are used for 
$Q_7$ and $Q_8$. To check the convergence of perturbation
theory at the intermediate scale $\mu_i\sim 1.5$ GeV
we split up the contributions from each operator into these 
two contributions. We also separate the NNLO corrections from the 
Wilson coefficients in the effective weak Hamiltonian separate,
labeling them with a [w].    For these three sources of
NNLO corrections,  in units of $1/A_{\rm v}^{\rm LO}$, we have 
\begin{eqnarray}
&&Q_1 :   (0.023+0.046 i)\,[\alpha_s] +\bigg( (-0.001 -0.002i)
\,[{\rm w}]+
(0.005+0.006i) \,[{\rm jet}] \bigg)\,[\alpha_s^2],
 \nonumber \\
&&Q_7 :  0.074 \,[\alpha_s]
+\bigg(-0.002\,[{\rm w}]+
0.015 \,[{\rm jet}] + 0.012 \,[{\rm hard}]\bigg)\,[\alpha_s^2], \nonumber \\
&&Q_8 :  0.01 \,[\alpha_s] +\bigg(0.001\,[{\rm w}]+ 0.001\,[{\rm jet}] + 
(0.001+0.005i)\,[{\rm hard}] \bigg)\,[\alpha_s^2].
\end{eqnarray}
In none of the cases is the correction at the jet scale 
$\mu_i = 1.5$~GeV unusually large.

\subsection{The SCET soft function}
\label{sec:SoftFunction}
In this subsection we explain our method for determining
the SCET soft function $\zeta_{V_\perp}$.  
We fix it by requiring that the matrix element of $Q_7$ 
is proportional to the tensor QCD form factor
$F^{B\to V_\perp}$ (often referred to as $T_1$).  
Using the SCET factorization formula for $Q_7$ 
we find 
\begin{equation}
\label{eq:Form}
 F^{B\to V_\perp}=
\frac{\Delta_7 C^{A}}{\Delta_7 C^{A(0)}}\zeta_{V_\perp}
- \frac{1}{\Delta_7 C^{B1(0)}}t_7^{{\rm II}}
\star \left(\frac{\sqrt{m_B} F}{4m_b} 
\phi_B \star f_{V_\perp} \phi_{V_{\perp}}\right).
\end{equation}
The recent LCSR-based update~\cite{Ball:2006eu} for the tensor 
QCD form factor yields $ F^{B \to K^*}=F^{B \to \phi}=0.31 \pm 0.04$ at
$\mu_{\rm QCD}=m_b$.  Inserting this into (\ref{eq:Form}) and 
treating the hard-spectator term as in the default scheme above
leads to  $\zeta_{{V}_\perp}(\mu=m_b) =0.35\pm 0.05$.  
This is considerably smaller than the value 
$\zeta_{{V}_\perp} \simeq 0.41$ used
in the SCET analysis in \cite{Becher:2005fg}, and it is mainly
for this reason that we find smaller branching fractions below.

We can use  this value for $\zeta_{V_\perp}$ to compare
the size of higher-order corrections to the factorization 
formula for the form factor.  We label the vertex term (v) and the 
hard spectator term (hs), and express each 
as an  expansion in $\alpha_s$.  Then the individual contributions
read 
\begin{eqnarray}
\frac{ F^{B\to V_\perp}}{\zeta_{V_\perp}}&=&
\big(1-0.15 [\alpha_s]-0.06 [\alpha_s^2]\big)[{\rm v}]+
\big(0.07 [\alpha_s]+0.03[\alpha_s^2]\big)[{\rm hs}] \,.
\end{eqnarray}
For both the $\alpha_s$ and $\alpha_s^2$ corrections the vertex
term is about twice as large as the hard-spectator term 
and comes with the opposite sign.

\subsection{Branching fractions}
\label{sec:BranchingFractions}

\begin{table}[t]
\caption{
Input parameters used in the calculation of ${\cal B}(B \to K^* \gamma)$
and ${\cal B}(B \to \phi \gamma)$. The  Gegenbauer
coefficients in the LCDAs are taken from the LCSR analysis reported in
\cite{Ball:2006eu}.}
\label{tab:input-param}
\begin{center}
\begin{tabular}{|l|l|}
\hline
Parameter & Value \\ \hline
$\alpha_s(m_Z)$                 & $0.1176$  \\
$V_{cs}^*V_{cb}$ & $-0.040\pm 0.002$ \\ 
$ \zeta_{{K^*}_\perp}(0)$ & $0.35 \pm 0.05$  \\
$ \zeta_{{\phi}_\perp} (0)$ & $0.35 \pm 0.05$  \\
$m_{b, {\rm pole}}$        & ($4.80 \pm 0.10$) GeV  \\
$m_{t, {\rm pole}}$        & ($171 \pm 2.0$) GeV\\
$\sqrt z = m_c / m_b$ & $0.27 \pm 0.06$ \\
$f_B$                      & ($205 \pm 25$) MeV  \\
$f_{B_s}$                      & ($240 \pm 30$) MeV  \\
$f_\perp^{(K^*)}$(1 GeV) & ($185 \pm 10$) MeV  \\
$f_\perp^{(\phi)}$(1 GeV) & ($186 \pm 9$) MeV  \\
$a_{\perp 1}^{(K^*)}$(1 GeV)  & $0.04 \pm 0.03$  \\
$a_{\perp 1}^{(\phi)}$(1 GeV)  & $0.0$  \\

$a_{\perp 2}^{(K^*)}$(1 GeV) & $0.15 \pm 0.10$ \\
$a_{\perp 2}^{(\phi)}$(1 GeV) & $0.20 \pm 0.20$  \\
$\lambda_B^{-1}$(1 GeV) & ($2.15 \pm 0.50$) GeV$^{-1}$  \\
$\sigma_B$(1 GeV)       & ($1.4 \pm 0.4$)  \\
\hline
\end{tabular}
\end{center}
\end{table}

We now convert our results for the amplitudes into estimates
for the branching fractions at NNLO. 
The most important uncertainties in the input parameters come
from $\zeta_{V_\perp}$, $\sqrt{z}=m_c/m_b$, $\lambda_B$, 
and the renormalization scales.  
To assess the uncertainty associated
with $\zeta_{V_\perp}$, $m_c$ and $\lambda_B$, 
we vary them in the ranges indicated in 
Table~\ref{tab:input-param}. 
The scale dependence of the branching fraction is completely
dominated by the vertex term. We treat this dependence as
explained in  Section \ref{sec:VertexAmplitude},
varying the scale $\mu_h=\mu_{\rm QCD}$ 
in the range  $m_b/\sqrt{2}<\mu_h<\sqrt{2}m_b $.
Including the corrections up to NNLO and discarding
terms of ${\cal O}(\alpha_s^3)$ and higher in the 
branching fractions, we find 
\begin{eqnarray}
{\cal B}(B^+ \to K^{*+}\gamma) &=& 
(4.6 \pm 1.2 \,[\zeta_{K^*}]  \pm 0.4\, [m_c]\pm 0.2 
\,[\lambda_B]\pm 0.1 \, [\mu] ) \times 10^{-5} ,\nonumber\\
{\cal B}(B^0 \to K^{*0}\gamma) &=& 
(4.3 \pm 1.1 \,  [\zeta_{K^*}] \pm 0.4 \,[m_c]  
\pm 0.2 \,[\lambda_B]\pm 0.1 \, [\mu])  
\times 10^{-5} ,\nonumber\\
{\cal B}(B_s \to \phi\gamma) &=& (4.3
\pm 1.1 \,  [\zeta_{\phi}] \pm 0.3 \,[m_c]
\pm 0.3 \,[\lambda_B] \pm 0.1 \, [\mu]  ) \times 10^{-5}.
\label{eq:brs}
\end{eqnarray}
In cases where the errors are asymmetric, we have taken the average
of the higher and lower values to get the symmetric form above.
The uncertainty in $|V_{cs}^* V_{cb}|$, which appears
as an overall factor multiplying the branching fractions,  
adds about a 10\% error to each decay mode.  
To obtain the branching fractions we used the following lifetimes 
(in units of $ps$)~\cite{HFAG}
\begin{equation}
\tau(B^0)=1.527 \pm 0.008;~~\tau(B^+)=1.643 \pm 0.010;
~~\tau(B_s) =1.451 \pm 0.028~.
\end{equation}
In addition to the lifetime differences, our analysis 
of the three decay modes includes differences in the meson 
decay constants, meson masses, and Gegenbauer moments of 
the light-meson LCDAs (we have assumed that SU(3) violating 
effects in the $B$-meson LCDAs are small).   
Other sources of isospin and SU(3) violation are not
included.  Concerning the $\phi$ and $K^*$ 
decay modes, the most important source of SU(3) violation 
is the difference between the SCET soft functions of the
two mesons. We discuss this in more
detail below, giving a result for the ratio of branching
fractions of these two decay modes.
A study of dynamical isospin breaking
contributions within QCD factorization was carried out in
\cite{Kagan:2001zk}.  From this study we expect the dynamical 
isospin violating effects to make only a small difference in 
the branching fractions.

It is important to keep in mind that we have not completed the NNLO
calculation for $Q_1$. The NNLO vertex correction
is only an estimate in the large-$\beta_0$ limit and the NNLO hard-spectator
correction related to $\Delta_1 C^{B1}$ is entirely absent.
To study the effects of possible deviations from  
large-$\beta_0$ limit we assign a 100\% uncertainty 
to the NNLO vertex correction from $Q_1$, evaluating 
the branching fractions using $2\Delta_1 C^{A(2)}$ and $\Delta_1 C^{A(2)}=0$.
For the hard spectator term we take the NNLO correction 
as $\pm 1$ its NLO value.  The corresponding uncertainties
in the branching fractions, to be added to the errors 
quoted in  (\ref{eq:brs}), are $\pm 0.5$ for the vertex corrections and
$\pm 0.1$ for the hard-spectator corrections.
The uncertainties associated with the unknown corrections 
to hard spectator scattering make little difference for 
the branching fraction. The uncertainties associated with the 
large-$\beta_0$ limit in the vertex term are rather large,
even though this is an ${\cal O}(\alpha_s^2)$ correction. 
We are very conservative  with the range in which we 
vary this correction, but even in the only existing calculation of 
NNLO corrections from $Q_1$ beyond the 
large-$\beta_0$ limit~\cite{Misiak:2006zs} for the inclusive case 
this is an issue.  In that paper the part of the $O(\alpha_s^2)$ correction 
to the matrix element of $Q_1$ beyond the large-$\beta_0$ limit 
(called $P_2^{(2) {\rm rem}}(z_0)$ in~\cite{Misiak:2006zs}) 
remains rather uncertain.

Adding together all the errors mentioned above in quadrature,
we obtain the final results for the branching fractions
\begin{eqnarray}
{\cal B}(B^+ \to K^{*+}\gamma) &=& 
(4.6 \pm 1.4 ) \times 10^{-5} ,\nonumber\\
{\cal B}(B^0 \to K^{*0}\gamma) &=& 
(4.3 \pm 1.4)  
\times 10^{-5} ,\nonumber\\
{\cal B}(B_s \to \phi\gamma) &=& (4.3
\pm 1.4) \times 10^{-5}.
\label{eq:finalbrs}
\end{eqnarray}
The NNLO estimates given in (\ref{eq:finalbrs}) are to be compared with
the experimental measurements summarized in Table 1.  We find
\begin{eqnarray}
\frac{{\cal B}(B^+ \to K^{*+}\gamma)_{\rm SM,NNLO}}
{{\cal B}(B^+ \to K^{*+}\gamma)_{\rm expt}}&=& 
1.1 \pm 0.35\, [\rm theory] \pm 0.07\, [\rm expt.] \,, \nonumber \\
\frac{{\cal B}(B^0 \to K^{*0}\gamma)_{\rm SM, NNLO}}
{{\cal B}(B^0 \to K^{*0}\gamma)_{\rm expt}}&=& 1.1 
 \pm 0.35\, [\rm theory] \pm 0.06 \, [\rm expt. ]\, , \nonumber \\
\frac{{\cal B}(B_s \to \phi \gamma)_{\rm SM, NNLO}}
{{\cal B}(B_s \to \phi\gamma)_{\rm expt}}&=& 0.8 
 \pm 0.2\, [\rm theory] \pm 0.3 \, [\rm expt. ]\, .
\label{eq:brsm-expt}
\end{eqnarray} 
Although the results are in reasonable  agreement with each other, 
the theory errors for the $B\to K^*\gamma$ decay modes 
are still much larger than the experimental ones.  
The dominant uncertainty is in the SCET soft function 
$\zeta_{V_\perp}$.  The remaining uncertainties 
would be greatly reduced by determining the  NNLO corrections
from $Q_1$ to the vertex term beyond the large-$\beta_0$ limit.
This would not only directly eliminate the uncertainty in
the NNLO correction to the hard-scattering kernel, 
it would also reduce the dependence on the charm-quark mass 
by fixing its perturbative definition.

Another measurement of interest is the ratio of the branching 
fractions of the $K^*$ and $\phi$ decay modes.  In the ratio,
only the errors in the quantities which are different for the
$B_s,\phi$ and $B,K^*$ mesons add significant uncertainties.
Since the spectator scattering amplitude is small compared
to the vertex term, to a good approximation the error is 
given by that in the ratio $\zeta_{K^*}/\zeta_{\phi}$. As 
an example, assuming $\zeta_{K^*}/\zeta_{\phi}=1\pm 0.1$,
we find for the ratio of branching fractions
\begin{equation}
\frac{{\cal B}(B^0 \to K^{*0}\gamma)}{{\cal B}(B_s \to \phi\gamma)} 
=1.0 \pm 0.2 \, .
\end{equation}
By comparison, the current experimental number is $0.7 \pm 0.3$.
Improved measurements of the $B_s \to \phi \gamma$ branching fraction, 
and a more accurate determination of the 
ratio of SCET soft functions, would allow for a comparison
between theory and experiment with smaller uncertainties 
than for the branching fractions themselves.

\section{Conclusions}
\label{sec:Conclusions}
We computed NNLO corrections to the hard-scattering kernels
entering the QCD factorization formula
for $B\to V\gamma$ decays.  We used soft-collinear effective
theory to separate contributions between the hard and 
hard-collinear scales and to resum large logarithms
depending on their ratio.
For the operators $Q_7$ and $Q_8$ we obtained exact expressions for 
the hard-scattering kernels for the vertex and hard spectator
corrections  up to NNLO.  The results 
for the vertex corrections provide an explicit demonstration
of factorization at two loops.
For the operator $Q_1$, we estimated its contribution to 
the vertex correction at NNLO using the large-$\beta_0$
limit. Its complete NNLO correction from hard spectator scattering 
was not obtained, but its contribution at the jet scale
was evaluated numerically and found to be small. 

As an application of our results we provided estimates of the
branching fractions for $B\to K^*\gamma$ and $B_s\to \phi\gamma$ 
decays at NNLO. The branching fractions
are very sensitive to the value of the SCET soft 
function $\zeta_{V_\perp}$.  We used updated results from 
QCD sum rules for the tensor form factor
$F^{B\to V_\perp}$ along with our NNLO results
for $Q_7$ to find $\zeta_{V_\perp} \simeq 0.35$.   
Since this value is considerably lower than the 
default value $\zeta_{V_\perp} \simeq 0.41$ used in 
the previous SCET analysis in \cite{Becher:2005fg},
we also find lower branching fractions.  
Our results for the $B\to K^*\gamma$ modes
show good agreement with the experimental data, but
the theory errors are still much larger than the experimental 
ones. Our result for $B_s \to \phi \gamma$, which has a comparable 
theoretical error as in the  $B\to K^*\gamma$ modes, is also in agreement
with the data within the large experimental error. 
The main theoretical uncertainty is in $\zeta_{V_\perp}$, which 
can be reduced by improved lattice or QCD sum-rule calculations.
On the perturbative
side, by far the most important issue is the calculation of 
the NNLO vertex correction for $Q_1$ beyond the large-$\beta_0$ limit.   
This requires the same diagrammatic calculation as 
the virtual corrections to inclusive $B\to X_s\gamma$, which remains
to be done.  Our results are also relevant 
for $B\to \rho\gamma$ and  $B\to \omega\gamma$, but for these 
decays a complete description also requires the perturbative 
corrections to weak annihilation, a topic we leave for future work.

{\bf Acknowledgments}: We would like to thank Alexander Parkhomenko for
collaboration in the early stages of this work, Guo-huai Zhu for 
correspondence on the numerical aspects of SCET resummation,
and Thomas Becher for useful discussions. One of us (C.G.) thanks
DESY for the hospitality in Hamburg where this work was carried out.
This work was supported in part by the
EU Contract No. MRTN-CT-2006-035482, FLAVIAnet.

\section{Appendix}
\subsection{Matrix elements }
In the section we give results for the UV renormalized on-shell
matrix elements $$\langle Q_i \rangle \equiv
\langle  q(p)\gamma(q)|Q_{i}|b(p_b)\rangle$$ 
in QCD.   The results given below are calculated in the 
$\MSbar$ renormalization scheme with 
$n_f=n_h+n_l$ flavors.  
For $Q_7$ and $Q_8$ we write 
\begin{equation}
\langle Q_i \rangle= \langle Q_{7,tree} \rangle
\left[\delta_{i7}+ \frac{C_F\alpha_s}{4\pi}D_i^{(1)}
+ \left(\frac{\alpha_s}{4\pi}\right)^2
C_F\left(C_F D_{iF}^{(2)}+ C_A D_{iA}^{(2)} + n_l D_{iL}^{(2)}
+ n_h D_{iH}^{(2)}\right)\right].
\end{equation}
For $Q_7$ the results are  \cite{Blokland:2005uk,Asatrian:2006ph}
(recall $L=\ln \mu/m_b$)
\begin{eqnarray}
D_7^{(1)}&=&-\frac{1}{\eps^2}-\frac{2L+2.5}{\eps}
-2 L^2-7L -6.8225  \\ &&
-\eps(1.3333 L^3 + 7 L^2 + 13.6449 L +13.4779) \nonumber \\
&&-\eps^2(0.6667 L^4+ 4.6667 L^3 + 13.6449 L^2 + 26.9559 L + 26.1412), 
\nonumber
\\
D_{7F}^{(2)}&=&\frac{0.5}{\eps^4}+\frac{2 L +2.5}{\eps^3}
+\frac{4L^2+12 L + 9.9475}{\eps^2}+
\frac{5.3333 L^3+26 L^2+44.7899 L+27.8816}{\eps}\nonumber \\
&&+5.3333 L^4 + 36 L^3+96.5798 L^2+144.1712 L+67.6519, \nonumber \\
D_{7A}^{(2)}&=&\frac{2.75}{\eps^3}+\frac{3.6667 L+3.5447}{\eps^2}
-\frac{4.1546 L+3.4386}{\eps} \nonumber \\
&&-4.8889 L^3-33.9758 L^2 -92.3415 L -83.8866, \nonumber \\
D_{7L}^{(2)}&=& -\frac{0.5}{\eps^3}-\frac{0.6667 L + 0.5556}{\eps^2}
+\frac{1.1111 L + 1.9799}{\eps} \nonumber \\
&&+0.8889 L^3+6.8889 L^2+19.9050 L+23.8254, \nonumber \\
D_{7H}^{(2)}&=&\frac{1.3333L}{\eps^2}+\frac{4 L^2+3.3333 L+0.5483}{\eps}
+6.2222 L^3+11.3333 L^2 + 14.1788 L + 0.2934, \nonumber
\end{eqnarray}
and for  $Q_8$ we have \cite{greub_prep}
\begin{eqnarray}
D_{8}^{(1)}&=&2.6667 L + 1.4734 + 2.0944 i
+\eps[2.6667 L^2+ 2.9468 L-1.1947+ i(4.1888L + 4.1888)] \nonumber \\
&&+\eps^2[1.7778L^3+2.9468 L^2-2.3894 L -5.5373 
+ i(4.1888 L^2+8.3776 L +2.1627)], \nonumber \\
D_{8F}^{(2)}&=&
D_8^{(1)}\left(-\frac{1}{\eps^2}-\frac{2L+2.5}{\eps}\right)-5.3333 L^3
-32.2802 L^2-50.9612 L -1.8875  \nonumber \\
&&-i(4.1888 L^2+31.4159 L +29.8299), \nonumber \\
D_{8A}^{(2)}&=&15.111 L^2+31.6617 L + 2.38332 + i(23.7365 L+28.0745), 
\nonumber\\
D_{8L}^{(2)}&=& -1.7778 L^2-4.0386 L -1.7170 -i(2.7925 L +4.4215),\nonumber \\
D_{8H}^{(2)}&=& -1.7778 L^2-4.0386 L +0.8829 -i 2.7925 L.
\end{eqnarray}

\subsection{The coefficients $\Delta_1 C^{B1(0)}$, $\Delta_7 C^{B1(1)}$ and 
$j_\perp^{(1)}$ }
Here we list the coefficients needed for the numerical analysis of 
spectator scattering which are 
not written in main text.  The lowest order expression for 
$\Delta_1 C^{B1}$ is  \cite{Becher:2005fg}
\begin{equation}
\Delta_1 C^{B1(0)}(u)= \frac{E_\gamma}{4\pi^2} \frac{2e}{3}
f\left(\frac{m_c^2}{ \bar u m_b^2}\right),
\end{equation}
where 
\begin{eqnarray}
f(x)&=&\theta\left(\frac{1}{4}-x\right)\left[1+4x
\left({\rm arctanh}(\sqrt{1-4x})-i \frac{\pi}{2}\right)^2\right]
\nonumber \\ &&
+\theta\left(x-\frac{1}{4}\right)
\left[1-4 x\left({\rm arctan}^2\frac{1}{\sqrt{4 x-1}}\right)\right].
\end{eqnarray}
For $Q_7$ the tree-level coefficient was given
in (\ref{eq:leadingC7}).  The one-loop correction is \cite{Beneke:2004rc,Becher:2004kk}
\begin{eqnarray}
\frac{\Delta_7 C^{B1(1)}}{\Delta_7 C^{B1(0)}}&=
& \frac{C_F\alpha_s}{4\pi}\frac{1}{2}
\bigg\{-4 \ln^2\frac{\mu}{m_b} -
2 \ln \frac{\mu}{m_b} -4 \ln\frac{\mu_{\rm QCD}}{m_b}
-\frac{\pi^2}{6}-\frac{4}{\Unew}\ln \bar u-2 
\nonumber \\ &&
+\frac{4\bar u}{\Unew}\left[\left(-2 \ln\frac{\mu}{m_b} -1\right)\ln \bar u+\ln^2 \bar u+
{\rm Li}_2(\Unew)\right]\bigg\} \nonumber \\ &&
+\frac{1}{2}\left(C_F-\frac{C_A}{2}\right)\frac{\alpha_s}{4\pi} \bigg\{
-\frac{4 \bar u}{\Unew}\left[\left(-2 \ln\frac{\mu}{m_b} -1\right)\ln \bar u+\ln^2 \bar u 
+{\rm Li}_2(\Unew)\right]
\nonumber \\ &&
-\frac{4(2- u)}{\bar u}\left[\left(-2 \ln \frac{\mu}{m_b} -1\right)
\ln \Unew+\ln^2\Unew
+{\rm Li}_2( \bar u)\right] \nonumber \\ &&
+\frac{4}{\bar u\Unew}{\rm Li}_2(\bar u)+\frac{4}{\Unew}
\left[{\rm Li}_2(\Unew)-\frac{\pi^2}{6}\right]
+4 \ln \bar u- 4 \ln \Unew -4\bigg\}.
\end{eqnarray}
The one-loop correction to the jet function can be obtained from, e.g.,
eq. (79) of \cite{Beneke:2005gs} after appropriate 
replacements. Calling the one-loop correction defined in eq. (79) 
of  \cite{Beneke:2005gs} $j_\perp^{\rm{BY}}$ after the authors of
that paper, we have
\begin{equation}
j_\perp^{(1)}(\tau,u,\omega)=-\frac{4\pi C_F\alpha_s}{N_c}
 \frac{1}{m_b \omega \bar u}\delta(\tau-u)
\left[\frac{\alpha_s}{4\pi} \, j_\perp^{\rm{BY}}(\bar \tau;u,\omega)\right].
\end{equation}

\subsection{RG functions}
Here we summarize the perturbative solutions to the RG exponents
in (\ref{eq:CAev}, \ref{eq:resummedCB}).  We define the expansion 
coefficients of the anomalous dimensions and the $\beta$-function
as 
\begin{eqnarray}
   \Gamma_{\rm cusp}(\alpha_s) &=& \Gamma_0\,\frac{\alpha_s}{4\pi}
    + \Gamma_1 \left( \frac{\alpha_s}{4\pi} \right)^2
    + \Gamma_2 \left( \frac{\alpha_s}{4\pi} \right)^3 + \dots \,,
    \nonumber\\
   \beta(\alpha_s) &=& -2\alpha_s \left[ \beta_0\,\frac{\alpha_s}{4\pi}
    + \beta_1 \left( \frac{\alpha_s}{4\pi} \right)^2
    + \beta_2 \left( \frac{\alpha_s}{4\pi} \right)^3 + \dots \right] ,
\end{eqnarray}
and similarly for the anomalous dimension $\gamma_J$. In terms of these 
quantities, the function $a$ (and $a_J$ with obvious replacements) is given by
\begin{eqnarray}\label{asol}
   a(\nu,\mu)
   &=& -\frac{\Gamma_0}{2\beta_0}\,\Bigg[
    \ln\frac{\alpha_s(\mu)}{\alpha_s(\nu)}
    + \left( \frac{\Gamma_1}{\Gamma_0} - \frac{\beta_1}{\beta_0} \right)
    \frac{\alpha_s(\mu) - \alpha_s(\nu)}{4\pi} \Bigg].
   \end{eqnarray}
The result for the Sudakov factor $S$ to this same order is 
\begin{eqnarray}
   S(\nu,\mu) &=& \frac{\Gamma_0}{4\beta_0^2}\,\Bigg\{
    \frac{4\pi}{\alpha_s(\nu)} \left( 1 - \frac{1}{r} - \ln r \right)
    + \left( \frac{\Gamma_1}{\Gamma_0} - \frac{\beta_1}{\beta_0}
    \right) (1-r+\ln r) + \frac{\beta_1}{2\beta_0} \ln^2 r \nonumber\\
   &&\mbox{}+ \frac{\alpha_s(\nu)}{4\pi} \Bigg[ 
    \left( \frac{\beta_1\Gamma_1}{\beta_0\Gamma_0} - \frac{\beta_2}{\beta_0} 
    \right) (1-r+r\ln r)
    + \left( \frac{\beta_1^2}{\beta_0^2} - \frac{\beta_2}{\beta_0} \right)
    (1-r)\ln r \nonumber\\
   &&\hspace{1.0cm}
    \mbox{}- \left( \frac{\beta_1^2}{\beta_0^2} - \frac{\beta_2}{\beta_0}
    - \frac{\beta_1\Gamma_1}{\beta_0\Gamma_0} + \frac{\Gamma_2}{\Gamma_0}
    \right) \frac{(1-r)^2}{2} \Bigg] \Bigg\},
\end{eqnarray}
where $r=\alpha_s(\mu)/\alpha_s(\nu)$. The cusp anomalous dimension
to three loops is 
\begin{eqnarray}
   \Gamma_0 &=& 4 C_F  \,, \nonumber\\
   \Gamma_1 &=& 4 C_F \left[ \left( \frac{67}{9} - \frac{\pi^2}{3} \right)
    C_A - \frac{20}{9}\,T_F n_f \right] 
 \,, \nonumber\\
   \Gamma_2 &=& 4 C_F \Bigg[ C_A^2 \left( \frac{245}{6} - \frac{134\pi^2}{27}
    + \frac{11\pi^4}{45} + \frac{22}{3}\,\zeta_3 \right) 
    + C_A T_F n_f  \left( - \frac{418}{27} + \frac{40\pi^2}{27}
    - \frac{56}{3}\,\zeta_3 \right) \nonumber\\
   &&\mbox{}+ C_F T_F n_f \left( - \frac{55}{3} + 16\zeta_3 \right) 
    - \frac{16}{27}\,T_F^2 n_f^2 \Bigg],
\end{eqnarray}
and the QCD $\beta$ function is 
\begin{eqnarray}
   \beta_0 &=& \frac{11}{3}\,C_A - \frac43\,T_F n_f  \,, \nonumber\\
   \beta_1 &=& \frac{34}{3}\,C_A^2 - \frac{20}{3}\,C_A T_F n_f
    - 4 C_F T_F n_f \,, \\
   \beta_2 &=& \frac{2857}{54}\,C_A^3 + \left( 2 C_F^2
    - \frac{205}{9}\,C_F C_A - \frac{1415}{27}\,C_A^2 \right) T_F n_f
    + \left( \frac{44}{9}\,C_F + \frac{158}{27}\,C_A \right) T_F^2 n_f^2.
\nonumber
    \end{eqnarray}

\subsection{Separating scales in the $\Delta_iC^{A}$ coefficients}
Here we list the NNLO coefficients  $\Delta_iC^{A}$ in the case
where we distinguish $L_{\rm QCD}$ from $L$.  
This is achieved by solving the RG equation (\ref{eq:ADdef}) perturbatively,
given the form (\ref{eq:JAD}) for the anomalous dimension $\gamma^A$.  
This has been done in \cite{Neubert:2005nt} and we can use those results after making 
appropriate replacements.  We find
\begin{eqnarray}
\Delta_7 C^{A(2)} &=& C_F^2\big[ 2L^4 
+ 10 L^3 + 4 L^2 L_{\rm QCD} +
26.1449 L^2 + 10 L L_{\rm QCD}+ 2 L_{\rm QCD}^2 \nonumber \\
&& +23.5022 L + 32.6449 L_{\rm QCD}
 +7.8159\big] \nonumber \\
&&+ C_F C_A\big[-4.8889 L^3 - 26.6425 L^2 -14.6667 L L_{\rm QCD}
+ 7.33333 L_{\rm QCD}^2 \nonumber \\
&& -63.7859 L - 28.5556  L_{\rm QCD} - 83.8866\big]
\nonumber \\
&& +C_F n_l \big[
0.8889 L^3+ 5.5556 L^2 + 2.6667 L L_{\rm QCD}
-1.3333 L_{\rm QCD}^2 \nonumber \\
&& + 17.0161 L + 2.8889 L_{\rm QCD} + 23.8254\big]
\nonumber \\
&&+ C_F n_h \left(-1.3333 L_{\rm QCD}^2+2.8889 L_{\rm QCD} - 0.810288 \right) \,,
\end{eqnarray}
\begin{eqnarray}
\Delta_8 C^{A(2)}&=&-C_F^2
\big[5.3333 L^2 L_{\rm QCD}+ 2.9468 L^2 + 13.333 L L_{\rm QCD}
+16 L_{\rm QCD}^2  \nonumber \\
&& + 7.3671 L + 43.5941 L_{\rm QCD} + 1.8875 \nonumber  \\
&&+ i (4.1888 L^2 + 10.4720 L+20.9440 L_{\rm QCD} +29.8299)\big]
\nonumber \\
&&+  C_F C_A \big[19.5556 L L_{\rm QCD}-4.4444 L_{\rm QCD}^2 
+10.8051 L + 20.8566 L_{\rm QCD}+ 2.3833 \nonumber \\ &&
   + i (15.3589 L + 8.3776 L_{\rm QCD} +  28.0745)\big]\nonumber \\
&&- C_F n_l \big[3.5556 L L_{\rm QCD}-1.7778 L_{\rm QCD}^2 
+ 1.9646 L + 2.0741 L_{\rm QCD} +1.7170 \nonumber \\ &&
+ i( 2.7925 L +4.4215)\big] \nonumber \\
&&+ C_F n_h \big[1.7778 L_{\rm QCD}^2 -2.0741 L_{\rm QCD} +0.8829 \big] \,,
\end{eqnarray}

\begin{eqnarray}
\Delta_1 C^{A(2)}&=& - \frac{3\beta_0}{2}\frac{m_b}{{\overline m}_b}C_F
\big[2.4691 L_{\rm QCD}^2 + l^{(2)} (z) L_{\rm QCD} 
+ r^{(2)}(z)\big] - 2 \beta_0 L_{\rm QCD}\Delta_1 C^{A(1)}\nonumber \\
&& +  2 \beta_0 L \Delta_1 C^{A(1)}\,.
\end{eqnarray}

\end{document}